\documentclass[%
superscriptaddress,
 amsmath,amssymb,
 superscriptaddress,
 aps,
 pre,
 twocolumn
]{revtex4-2}
\usepackage{graphicx}%
\usepackage{dcolumn}%
\usepackage{bm}%
\usepackage{xcolor}
\usepackage{float}
\usepackage{appendix}
\usepackage{multirow}

\begin{document}
\title{Local and global measures of the shear moduli of jammed disk packings}

\author{Shiyun Zhang}
\thanks{These authors contributed equally and are listed in alphabetical order.}
\affiliation{Department of Physics, University of Science and Technology of China, Hefei, Anhui 230026, China}
\affiliation{Department of Mechanical Engineering and Materials Science, Yale University, New Haven, Connecticut 06520, USA}

\author{Weiwei Jin}
\thanks{These authors contributed equally and are listed in alphabetical order.}
\affiliation{Department of Mechanical Engineering and Materials Science, Yale University, New Haven, Connecticut 06520, USA}

\author{Dong Wang}
\affiliation{Department of Mechanical Engineering and Materials Science, Yale University, New Haven, Connecticut 06520, USA}

\author{Ding Xu}
\affiliation{Department of Physics, University of Science and Technology of China, Hefei, Anhui 230026, China}

\author{Jerry Zhang}
\affiliation{Department of Mechanical Engineering and Materials Science, Yale University, New Haven, Connecticut 06520, USA}

\author{Mark D. Shattuck}
\affiliation{Benjamin Levich Institute and Physics Department, The City College of New York, New York, New York 10031, USA}

\author{Corey S. O'Hern}
\affiliation{Department of Mechanical Engineering and Materials Science, Yale University, New Haven, Connecticut 06520, USA}
\affiliation{Department of Physics, Yale University, New Haven, Connecticut 06520, USA}
\affiliation{Department of Applied Physics, Yale University, New Haven, Connecticut 06520, USA}
\affiliation{Graduate Program in Computational Biology and Bioinformatics, Yale University, New Haven, Connecticut 06520, USA}

\begin{abstract}
Strain-controlled isotropic compression gives rise to jammed packings of repulsive, frictionless disks with either positive or negative global shear moduli.  We carry out computational studies to understand the contributions of the negative shear moduli to the mechanical response of jammed disk packings.  We first decompose the ensemble-averaged, global shear modulus as $\langle G\rangle = (1-{\cal F}_-) \langle G_+ \rangle + {\cal F}_- \langle G_-\rangle$, where ${\cal F}_-$ is the fraction of jammed packings with negative shear moduli and $\langle G_+\rangle$ and $\langle G_-\rangle$ are the average values from packings with positive and negative moduli, respectively.  We show that $\langle G_+\rangle$ and $\langle|G_-|\rangle$ obey different power-law scaling relations above and below $pN^2 \sim 1$. For $pN^2 > 1$, both $\langle G_+\rangle N$ and $\langle|G_-|\rangle N \sim  (pN^2)^{\beta}$, where $\beta \sim 0.5$ for repulsive linear spring interactions. Despite this, $\langle G\rangle N \sim (pN^2)^{\beta'}$ with $\beta' \gtrsim 0.5$ due to the contributions from packings with negative shear moduli. We show further that the probability distribution of global shear moduli ${\cal P}(G)$ collapses at fixed $pN^2$ and different values of $p$ and $N$. We calculate analytically that ${\cal P}(G)$ is a Gamma distribution in the $pN^2 \ll 1$ limit.  As $pN^2$ increases, the skewness of ${\cal P}(G)$ decreases and ${\cal P}(G)$ becomes a skew-normal distribution with negative skewness in the $pN^2 \gg 1$ limit.  We also partition jammed disk packings into subsystems using Delanunay triangulation of the disk centers to calculate local shear moduli. We show that the local shear moduli defined from groups of adjacent triangles can be negative even when $G > 0$.  The spatial correlation function of local shear moduli $C({\vec r})$ displays weak correlations for $pn_{\rm sub}^2 < 10^{-2}$, where $n_{\rm sub}$ is the number of particles within each subsystem. However, $C({\vec r})$ begins to develop long-ranged spatial correlations with four-fold angular symmetry for $pn_{\rm sub}^2 \gtrsim 10^{-2}$. 

\end{abstract}
\maketitle

\section{Introduction}
\label{intro}

Particulate materials, such as packings of bubbles~\cite{katgert2013jamming}, droplets~\cite{clusel2009granocentric}, colloids~\cite{pradeep2021}, and grains~\cite{zhang2010jamming}, jam into a solid-like state when they are compressed above jamming onset, while the internal structure remains disordered. A distinguishing feature of jammed solids is that they possess a nonzero shear modulus $G$, in addition to a nonzero bulk modulus $B$~\cite{o2003jamming}. Numerous computational and theoretical studies have employed the frictionless, soft-particle model~\cite{zhang2005jamming,xu2005random,silbert2009normal,jin2020}, which assumes pairwise, purely repulsive interactions between spherical particles, to study the onset of jamming in particulate materials.  Prior results have shown that at high pressures the shear modulus for jammed packings of spherical particles scales as a power law, $G \sim p^{\beta}$, where the scaling exponent $\beta$ depends on the form of the purely repulsive interaction potential, but not on the spatial dimension~\cite{o2003jamming,goodrich2014,zamponi_G_theory}.   

In most prior studies of jammed packings of frictionless, soft particles, packings are generated by isotropically compressing a collection of particles when the shape of the bounding box is fixed.  In this ``compression-only'' protocol, the shear modulus of a given packing can be negative and the boundaries of the system provide the necessary shear stress to prevent particles from flowing~\cite{dagoisbohy2012,goodrich2014,wang2021shear}. 
In contrast, a shear-stabilized packing protocol was proposed to generate jammed systems that are stable to shear in all directions by allowing all degrees of freedom of the boundary to change during energy minimization~\cite{dagoisbohy2012}. 
The two different protocols generate packings with different mechanical properties, resulting in the question of whether jammed packings with negative shear moduli should be excluded from the ensemble when using the ensemble average to represent the shear modulus in the large-system limit~\cite{dagoisbohy2012,goodrich2012,goodrich2014,wang2021shear,vanDeen}. 

In previous studies of jammed packings generated by the compression-only protocol, we showed that the pressure-dependent shear modulus has two contributions~\cite{vanderwerf2020pressure}: 1) continuous variations in the shear modulus with pressure from geometrical families, and 2) discontinuous jumps in the shear modulus from changes in the interparticle contact network. Geometrical families correspond to jammed packings at different pressures that are related to each other with the {\it same} interparticle contact network. For purely repulsive linear spring interactions~\cite{xu2006measurements}, the shear modulus of a near isostatic geometrical family can be approximated as $G/G_0 \sim 1 - p/p_0$, where $G_0$ is the shear modulus at $p=0$ and $p_0$ is the pressure at which $G=0$. From this form, it is clear that $G$ would become negative if particles in the jammed packing do not rearrange as the pressure increases. Hence, jammed packings with negative $G$ can be considered as natural members of the ensemble, which raises the question of how negative shear moduli affect the power-law scaling of the ensemble-averaged shear modulus $\langle G \rangle$. 

It is well-known that amorphous solids exhibit spatial heterogeneity at the particle scale~\cite{mizuno2013,mizuno2016,lemaitre2014Sigma_corr,gelin2016localG_corr,tong2020} in response to boundary-driven deformations.  Understanding these spatial heterogeneities is essential for linking bulk mechanical properties to particle-scale interactions and motion~\cite{wyart2, SoftSpots2011, Falk_tau2016, TongPsi2019,hu2022origin}. In particular, it has been shown that an affine deformation applied to an amorphous solid will give rise to strongly nonaffine particle-scale motion to restore force balance in the system~\cite{maloney,zaccone2011,richard20,jin2021}, which makes it more difficult to define local stress and strain for subdomains of amorphous solids. The strongly inhomogeneous stress and strain are believed to play a central role in controlling the anomalous acoustic excitations and bulk mechanical properties of amorphous solids~\cite{mizuno, LocalG_BP2007,marruzzo2013heterogeneous,mizuno2014acoustic,zaccone2011,zaccone2014,cui2019,baggioli2021}. However, despite its importance, it is not clear which definitions of local stress and strain best characterize their local structural and mechanical properties and which should be used to connect the local to the global mechanical response ~\cite{tsamados2009localG_size,localG_2004,mizuno2013,mizuno2016,gelin2016localG_corr}.  

In this work, we carry out computational studies to generate jammed binary disk packings (interacting via repulsive linear spring forces) using isotropic compression, while controlling the shape of the confining box. We focus on the mechanical response of jammed disk packings to applied simple shear and characterize the distribution of the global shear moduli (including both positive and negative values) as a function of the pressure $p$ and system size $N$. We also develop a novel method to calculate the local shear moduli $g$ of jammed disk packings as a function of the size of the subsystem $n_{\rm sub}$, and compare these results to those using other methods. 

We find several key results. First, we show that the separate contributions $\langle G_+\rangle$ and $\langle G_-\rangle$ to the ensemble-averaged shear modulus, $\langle G\rangle = (1-{\cal F}_-)\langle G_+\rangle +{\cal F}_- \langle G_-\rangle$, where ${\cal F}_-$ is the fraction of jammed packings with $G<0$, obey different scaling relations with pressure $p$ above and below $pN^2 \sim 1$. For $pN^2 < 1$, $\langle|G_-|\rangle N \sim pN^2$ and $(\langle G_+\rangle-G_0/(1-{\cal F}_-))N \sim (pN^2)^{\eta_+}$, where $G_0 \sim N^{-1}$ and $\eta_+ \sim 1.33$. In contrast, for $pN^2 > 1$, both $\langle G_+\rangle$ and $\langle|G_-|\rangle \sim p^{\beta}$, where $\beta \sim 0.5$. We find that the power-law scaling exponent $\beta \gtrsim 0.5$ for the ensemble-averaged shear modulus $\langle G \rangle$ since the fraction of packings with negative shear moduli decreases strongly with increasing $p$ for $pN^2 >1$. Second, we show analytically that the form for the probability distribution ${\cal P}(G)$ in the $pN^2 \rightarrow 0$ limit becomes a Gamma distribution with shape parameter $k=0.5$.  In contrast, when $pN^2 \gg 1$, ${\cal P}(G)$ becomes a left-skewed Gaussian distribution. 
Third, using a Delaunay triangulation method for calculating the local shear modulus $g$, we show that the shear modulus for single triangles, whose vertices represent the centers of three nearest neighbor disks, 
decreases linearly with pressure $g_t \sim g_0 -\lambda p$, where $g_0$ and the coefficient $\lambda$ depend on the triangle's orientation. This result is consistent with the dependence of the global shear modulus with pressure for jammed packings within geometrical families. Further, there can be an abundance of negative local shear moduli of subsystems composed of Delaunay triangles even for jammed packings with $G>0$. We find only weak spatial correlations in $g$ over a wide range of $pn_{\rm sub}^2 < 10^{-2}$, where $n_{\rm sub}$ is the subsystem size. In contrast, local shear moduli calculated by assuming that the local strain tensor is affine possess long-ranged spatial correlations with four-fold angular symmetry for all values of $pn_{\rm sub}^2$. These results elucidate the influence of negative shear moduli on the ensemble-averaged mechanical properties of jammed disk packings and provide promising directions for linking their local and global mechanical response. 

\begin{figure}[t]
    \centering
    \includegraphics[width=0.3\textwidth]{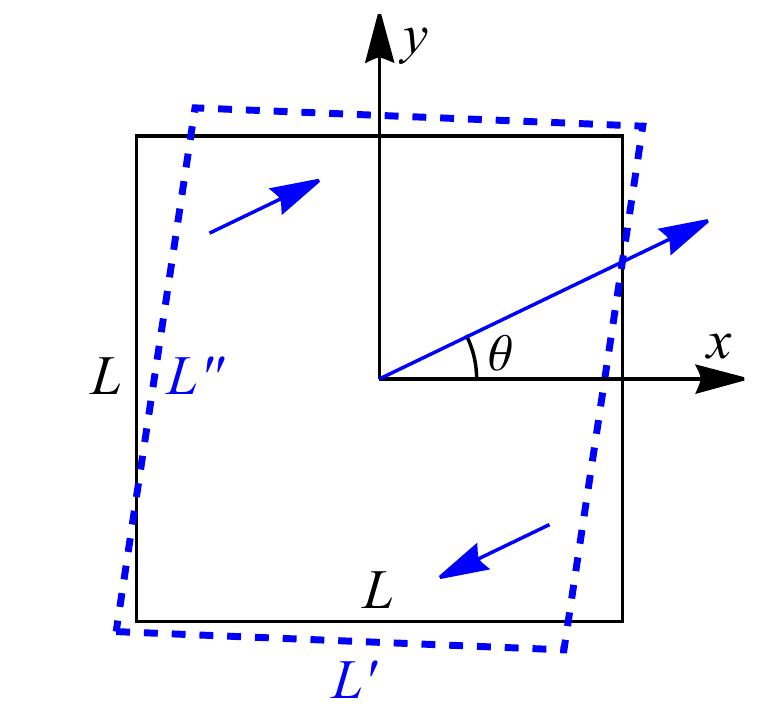}
    \caption{Sketch of a simple shear deformation (Eq.~\ref{simple_shear}) applied to a square cell (with side length $L$ and area $A=L^2$) at an angle $\theta$ to the $x$-axis. The sides of the undeformed square cell (black solid lines) are aligned with the $x$- and $y$-axes. The deformed cell (blue dashed lines) has area $A$ and side lengths $L'=\|{\overline F} (L,0)^{\rm T}\|$ and $L''=\|{\overline F} (0,L)^{\rm T}\|$.}
    \label{fig:shear}
\end{figure}

The remainder of the article is organized as follows.
In Sec.~\ref{methods}, we introduce the purely repulsive linear spring potential for modeling the interactions between disks, the protocol used to generate the jammed disk packings, and the methods to calculate their local and global shear moduli. We present our main results in Sec.~\ref{results} including the calculations of $\langle G \rangle$, $\langle G_+\rangle$, and $\langle |G_-|\rangle$ as a function of $p$ and $N$ and the probability distributions and spatial correlations of the local shear moduli (for different $n_{\rm sub}$) using the affine-strain and Delaunay triangulation methods. The conclusions and promising future research directions are provided in Sec.~\ref{secsum}. We also include three appendices. In Appendix~\ref{appxtristiff}, we derive the stiffness tensor for the five types of Delaunay triangles in binary disk packings in the low-pressure limit. In Appendix~\ref{appxdistG}, we provide additional data for ${\cal P}(G)$ at intermediate values of $pN^2$. In Appendix~\ref{appxPDF0}, we derive the form of ${\cal P}(G)$ for disk packings at jamming onset.

\section{Methods}
\label{methods}
\subsection{Model system and packing generation protocol}

We study the mechanical properties of jammed packings of $N$ frictionless disks with the same mass $m$ in two dimensions. We consider a range of system sizes, including $N=64$, $128$, $256$, and $1024$ to investigate the finite-size effects. The disks interact via the pairwise, purely repulsive linear spring potential,
\begin{equation}
    \label{potential}
    U(r_{ij}) = \frac{\epsilon}{2} \left( 1- \frac{r_{ij}}{\sigma_{ij}} \right)^{2}
    \Theta \left( 1- \frac{r_{ij}}{\sigma_{ij}} \right),
\end{equation}
where $\epsilon$ is the characteristic energy scale, $r_{ij}$ is the separation between the centers of disks $i$ and $j$, $\sigma_{ij} = (\sigma_i + \sigma_j)/2$ is the average of their diameters $\sigma_i$ and $\sigma_j$, and  $\Theta(\cdot)$ is the Heaviside step function. The total potential energy $U = \sum_{i>j} U(r_{ij})$ is obtained by summing $U(r_{ij})$ over all distinct disk pairs that are in contact. We focus on binary mixtures with $N/2$ large and $N/2$ small particles and the diameter ratio of the large to small disk, $\sigma_{l}/\sigma_{s}=1.4$, which inhibits crystallization~\cite{perera1998}. Below, we will display the data using $m$, $\sigma_s$, and $\epsilon$ as the units for mass, length, and energy, respectively. 

\begin{figure}[t]
    \centering
    \includegraphics[width=0.425\textwidth]{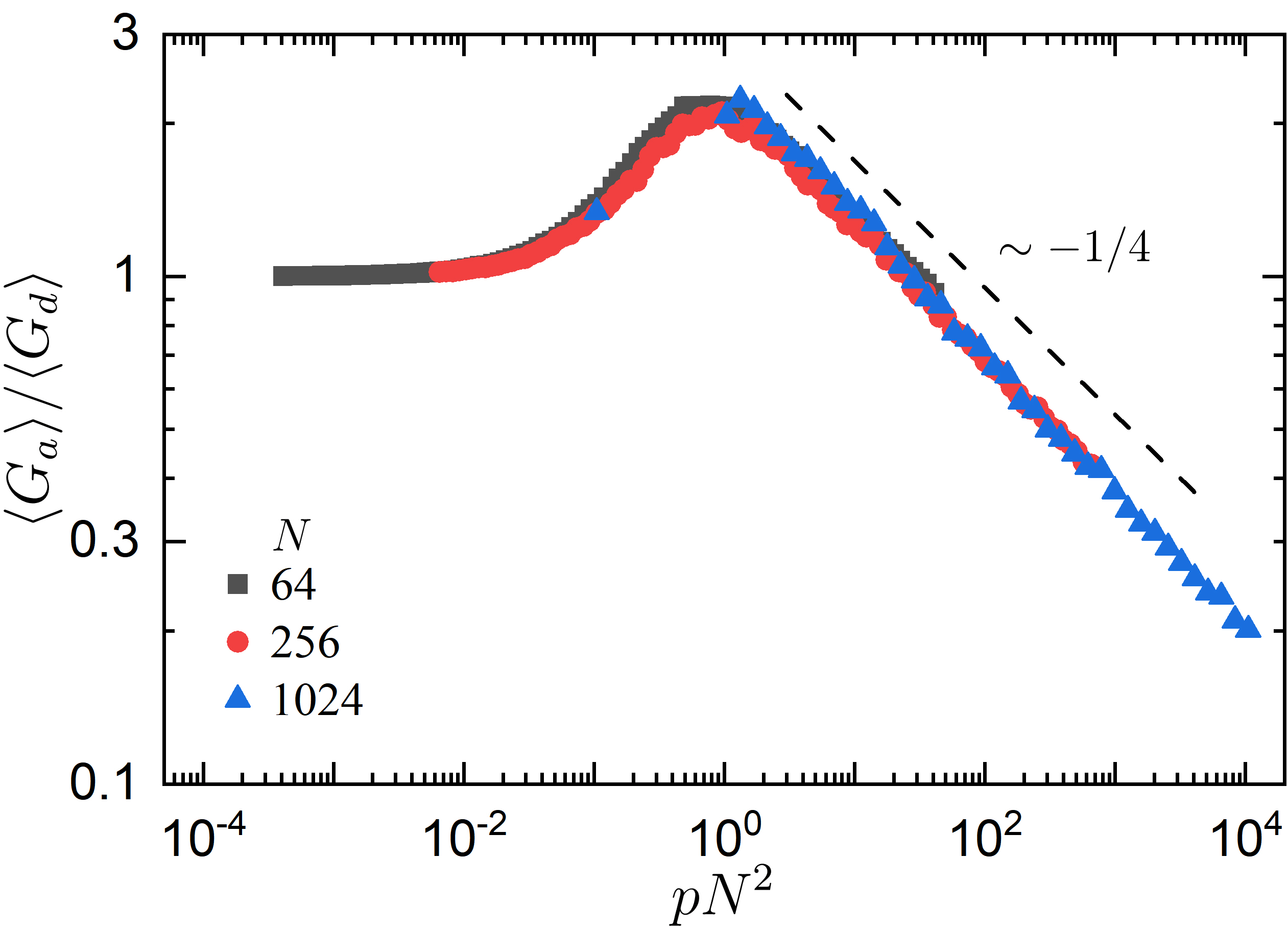}
    \caption{The ensemble-averaged amplitude of the shear modulus $\langle G_a \rangle$ (normalized by $\langle G_d \rangle$ in Eq.~\ref{eq:G}) plotted as a function of $pN^2$ for several system sizes $N=64$, $256$, and $1024$. The dashed line has a slope of $-0.25$. Similar results were found in Ref.~\cite{goodrich2014}.}
    \label{fig:aniso}
\end{figure}

To generate jammed packings, we first randomly placed $N$ disks in a square box with the side length $L$ and periodic boundary conditions in the $x$- and $y$-directions at initial packing fraction $\phi_0 =0.83$. We then perform minimization of the enthalpy $H=U+p' L^{2}$, where $p'$ is the target pressure~\cite{goodrich2014,minimizeH2014}, using the fast inertial relaxation engine (FIRE) minimization method~\cite{bitzek2006structural} with a fixed square box shape. 
The enthalpy minimization is terminated when the magnitude of the total force on each disk $i$ satisfies $|{\vec f}_i| < 10^{-14}$ and the pressure satisfies $|p-p'| < 10^{-14}$. 

The global stress tensor of each jammed disk packing is calculated via the virial expression:
\begin{equation}
\label{stress_tensor}
\Sigma_{\alpha \beta} = L^{-2} \sum^N_{i>j} r_{ij\alpha} f_{ij\beta},
\end{equation}
where $r_{ij\alpha}$ is the $\alpha$-component of the separation vector $\vec{r}_{ij}=(x_{ij},y_{ij})^{\rm T}$ pointing from the center of disk $j$ to the center of disk $i$ and $f_{ij\beta}$ is the $\beta$-component of the interparticle force $\vec{f}_{ij}=-(dU/dr_{ij}){\hat r}_{ij}$ on disk $i$ from $j$. The pressure and shear stress are defined as $p=( \Sigma_{xx}+\Sigma_{yy})/2$ and $\Sigma = -\Sigma_{xy}$.

We first generate an ensemble of ${\mathcal N}_c \sim 10^3$ jammed disk packings at low pressure $p=10^{-7}$. We then compress each of the packings in small pressure increments $\Delta p$ with each increment in pressure followed by enthalpy minimization. We choose $\Delta p$ such that we have $N_p  \approx 10^{3}$ pressure values evenly spaced on a logarithmic scale between $p = 10^{-7}$ and $10^{-2}$. 

\subsection{Calculation of global and local shear moduli}
\subsubsection{Global shear modulus}

We calculate the global shear modulus of each packing using the expression: $G=G^a-G^{na}$~\cite{maloney,zaccone2011}, where the affine term $G^{a}$ is the response to the applied affine simple shear strain and the nonaffine term $G^{na}$ gives the nonaffine response of the system as it relaxes to a new potential energy minimum after the applied simple shear. 
A simple shear increment $\delta \gamma$ applied to the packing at an angle $\theta$ to the $x$-axis, as illustrated in Fig.~\ref{fig:shear}, changes the position of disk $i$ to $(x_i^{\prime},y_i^{\prime})^{\rm T}={\overline F}(x_i^0,y_i^0)^{\rm T}$, where $(x_i^0,y_i^0)^{\rm T}$ is the original position of the disk and 
\begin{equation}
\label{simple_shear}
   {\overline F}=
   \begin{bmatrix}
1-\frac{1}{2} \delta\gamma  \sin 2\theta  & \frac{1}{2} \delta\gamma  \left(1+\cos 2\theta \right) \\
-\frac{1}{2} \delta\gamma  \left(1-\cos 2\theta \right) & 1+\frac{1}{2} \delta\gamma  \sin 2\theta 
\end{bmatrix}
\end{equation}
is the deformation gradient tensor. This deformation preserves the area of the box $A=L^2$, but changes the side lengths of the confining box to $L'=\|{\overline F} (L,0)^{\rm T}\|$ and $L''=\|{\overline F} (0,L)^{\rm T}\|$, where $\|(x,y)^{\rm T}\|=\sqrt{x^2+y^2}$. 
The affine and nonaffine contributions to the shear modulus are
\begin{subequations}
\begin{align}
    \label{Ga-Gna}
    G^a\;&=\frac{1}{L^2}\frac{\partial^2U}{\partial\gamma^2},\\
    G^{na} &= \frac{1}{L^2}\Xi_{i\alpha}M^{-1}_{i\alpha j\beta}\Xi_{j\beta},
    \end{align}
\end{subequations}
where $M_{i\alpha j\beta}=\frac{\partial^2 U}{\partial r_{i\alpha} \partial r_{j\beta}}$ is the dynamical matrix, $r_{i\alpha}$ is the $\alpha$-component of ${\vec r}_i=(x_i,y_i)^{\rm T}$, and ${\vec \Xi}_i=\frac{\partial^2 U}{\partial {\vec r}_i \partial \gamma}$ is the virtual force incurred after a small shear strain increment. 

Both the shear stress and shear modulus vary sinusoidally with the angle $\theta$ at which the simple shear strain is applied~\cite{dagoisbohy2012,goodrich2014}:
\begin{subequations}
\begin{align}
\label{eq:stress}
\Sigma&=\Sigma_{a}\sin2(\theta-\theta_S),\\
\label{eq:G}
G\;\:&=G_{a}\sin4(\theta-\theta_G)+G_{d},
\end{align}
\label{eq:stressG}
\end{subequations}

\noindent where $\Sigma_{a}$ and $G_{a}$ are the amplitudes of the shear stress and shear modulus, $\theta_S$ and $\theta_G$ are the phase shifts of the shear stress and shear modulus, and $G_{d}$ is the angle-averaged shear modulus. The ensemble-averaged amplitude of the shear modulus $\langle G_a \rangle$ (normalized by $\langle G_d \rangle$) is nonzero in the $pN^2 \rightarrow 0$ limit, whereas $\langle G_a \rangle/\langle G_d \rangle \sim 1/(pN^2)^{\eta}$ (with $\eta \sim 0.25$) tends to zero in the $pN^2 \gg 1$ limit~\cite{goodrich2014} as shown in Fig.~\ref{fig:aniso}.

\subsubsection{Local shear modulus}
\label{localg}
We employed two methods to calculate the local shear moduli $g$ of subsystems of jammed packings. In the first method, which assumes an affine response of each subsystem, each square system is divided into $n\times n$ smaller identical subsystems with an average of $n_{\rm sub} = N/n^2$ disks per subsystem. The local virial stress tensor for each subsystem $\ell$ is
\begin{equation}
    \label{localstress}
   \Sigma^\ell_{\alpha\beta} =\frac{n^2}{ L^{2}} \sum_{i>j} r_{ij\alpha} f_{ij\beta}  \frac{q_{ij}}{r_{ij}},
\end{equation}
where $q_{ij}$ is the length of the portion of $r_{ij}$ that is inside subsystem $\ell$. For this method, we assume that the imposed global strain represents the local strain of all subsystems. The local pressure and shear stress are defined as $p^{\ell}=( \Sigma^\ell_{xx}+\Sigma^\ell_{yy})/2$ and $\Sigma^\ell = -\Sigma^\ell_{xy}$. Thus, the local shear modulus is $g^\ell=d \Sigma^\ell /d \gamma$. The area-weighted sum over all subsystems of the local shear stress $\Sigma^\ell$ and local shear modulus $g^\ell$ yield the global shear stress $\Sigma$ and shear modulus $G$. 

In the second approach, we seek to more accurately characterize the local strain of each subsystem. We perform Delaunay triangulation using the disk centers as the vertices of the triangles, and define the stress and strain tensors for each Delaunay triangle. We first apply three types of deformations separately to a given jammed packing: 1) uniaxial compression in the $x$-direction (denoted as $\mathcal{D}_1$), 2) uniaxial compression in the $y$-direction ($\mathcal{D}_2$), and 3) simple shear with the $x$-axis as the shear direction and the $y$-axis as the shear gradient direction ($\mathcal{D}_3$). The deformation gradient tensors for these three boundary deformations are:  
\begin{subequations}
\label{affinef}
\begin{align}
&{\overline F}\mid_{\mathcal{D}_1}=\begin{bmatrix}
1-\varepsilon  & 0 \\
0 & 1 
\end{bmatrix},\\
&{\overline F}\mid_{\mathcal{D}_2}=\begin{bmatrix}
1  & 0 \\
0 & 1-\varepsilon 
\end{bmatrix},\\
&{\overline F}\mid_{\mathcal{D}_3}=\begin{bmatrix}
1  & \varepsilon \\
0 & 1 
\end{bmatrix},
\end{align}
\end{subequations}
where $\varepsilon$ is the strain amplitude of the affine deformation.
After imposing a given affine deformation to the packing (i.e. the boundary and disk positions), the disks are moved nonaffinely according to the nonaffine ``velocity''~\cite{maloney},
\begin{equation}
\frac{d {\vec r}_m}{d\varepsilon}=-M_{mn}^{-1}{\vec \Xi}_n
\label{eq:nonaff}
\end{equation}
with the boundary held fixed. Using the updated disk positions $(x_m^{\prime},y_m^{\prime})^{\rm T}={\overline F}(x_m^0,y_m^0)^{\rm T}+\varepsilon \frac{d {\vec r}_m}{d\varepsilon}$ from Eqs.~\ref{affinef} and~\ref{eq:nonaff}, we can calculate the deformation gradient tensor, 
\begin{equation}
{\overline F}_i^\Delta\mid_\mathcal{D}=
\begin{bmatrix}
 x_{12}^\prime & x_{13}^\prime \\
 y_{12}^\prime & y_{13}^\prime \\
\end{bmatrix}
\begin{bmatrix}
 x_{12}^0 & x_{13}^0 \\
 y_{12}^0 & y_{13}^0 \\
\end{bmatrix}^{-1}\bigg|_\mathcal{D},
\end{equation}
for each triangle $i$ (with vertex labels $1$, $2$, and $3$) in a jammed packing with a given applied deformation $\mathcal{D}$. Using ${\overline F}_i^\Delta\mid_\mathcal{D}$, we can determine the associated Green-Lagrangian strain tensor,
\begin{equation}
{\overline E}_i^{\Delta}\mid_{\mathcal{D}}=\frac{1}{2}(({\overline F}_i^\Delta)^{\rm T}{\overline F}_i^\Delta-{\overline I})\mid_{\mathcal{D}},
\label{eq:Etri}
\end{equation}
where ${\overline I}$ is the $2\times 2$ identity matrix, and the difference in the 2nd Piola-Kirchhoff material stress tensor for triangle $i$ before and after the deformation,
\begin{equation}
{\overline \Sigma}_i^{m,\Delta}\mid_\mathcal{D}=\det({\overline F}_i^\Delta)({\overline F}_i^\Delta)^{-1} {\overline \Sigma}_i^{\Delta} ({\overline F}_i^\Delta)^{-\rm T}\mid_\mathcal{D}-{\overline \Sigma}_{i}^{\Delta},
\label{eq:Smtltri}
\end{equation}
which are used to calculate the $3\times 3$ stiffness matrix of each triangle,  
\begin{equation}
{\hat C}_i^\Delta=
    \begin{bmatrix}
    c_{xxxx} & c_{xxyy}  & c_{xxxy}\\
    c_{yyxx} & c_{yyyy}  & c_{yyxy}\\
    c_{xyxx} & c_{xyyy}  & c_{xyxy}\\
    \end{bmatrix}.
    \label{eq:C}
\end{equation}
The nine components of ${\hat C}_i^\Delta$ can be obtained from Hooke's law relating stress and strain, i.e. by solving the following set of nine equations: 
\begin{subequations}
\begin{align}
  \begin{bmatrix}
  \Sigma_{ixx}^{m,\Delta}\\
  \Sigma_{iyy}^{m,\Delta}\\
  \Sigma_{ixy}^{m,\Delta}
  \end{bmatrix}_{\mathcal{D}_{1}}={\hat C}_i^\Delta
  \begin{bmatrix} 
  E_{ixx}^{\Delta}\\
  E_{iyy}^{\Delta}\\
  2E_{ixy}^{\Delta}
  \end{bmatrix}_{\mathcal{D}_{1}},\label{eq:stifftri1}\\
  \begin{bmatrix}
  \Sigma_{ixx}^{m,\Delta}\\
  \Sigma_{iyy}^{m,\Delta}\\
  \Sigma_{ixy}^{m,\Delta}
  \end{bmatrix}_{\mathcal{D}_{2}}={\hat C}_i^\Delta
  \begin{bmatrix} 
  E_{ixx}^{\Delta}\\
  E_{iyy}^{\Delta}\\
  2E_{ixy}^{\Delta}
  \end{bmatrix}_{\mathcal{D}_{2}},\label{eq:stifftri2}\\
  \begin{bmatrix}
  \Sigma_{ixx}^{m,\Delta}\\
  \Sigma_{iyy}^{m,\Delta}\\
  \Sigma_{ixy}^{m,\Delta}
  \end{bmatrix}_{\mathcal{D}_{3}}={\hat C}_i^\Delta
  \begin{bmatrix} 
  E_{ixx}^{\Delta}\\
  E_{iyy}^{\Delta}\\
  2E_{ixy}^{\Delta}
  \end{bmatrix}_{\mathcal{D}_{3}}.\label{eq:stifftri3}
\end{align}
\label{eq:stifftri}
\end{subequations}
In this work, since we are interested in studying the shear modulus, we focus on the component $c_{xyxy} \equiv g^\ell$. 

\begin{figure}[t]
    \centering
    \includegraphics[width=0.425\textwidth]{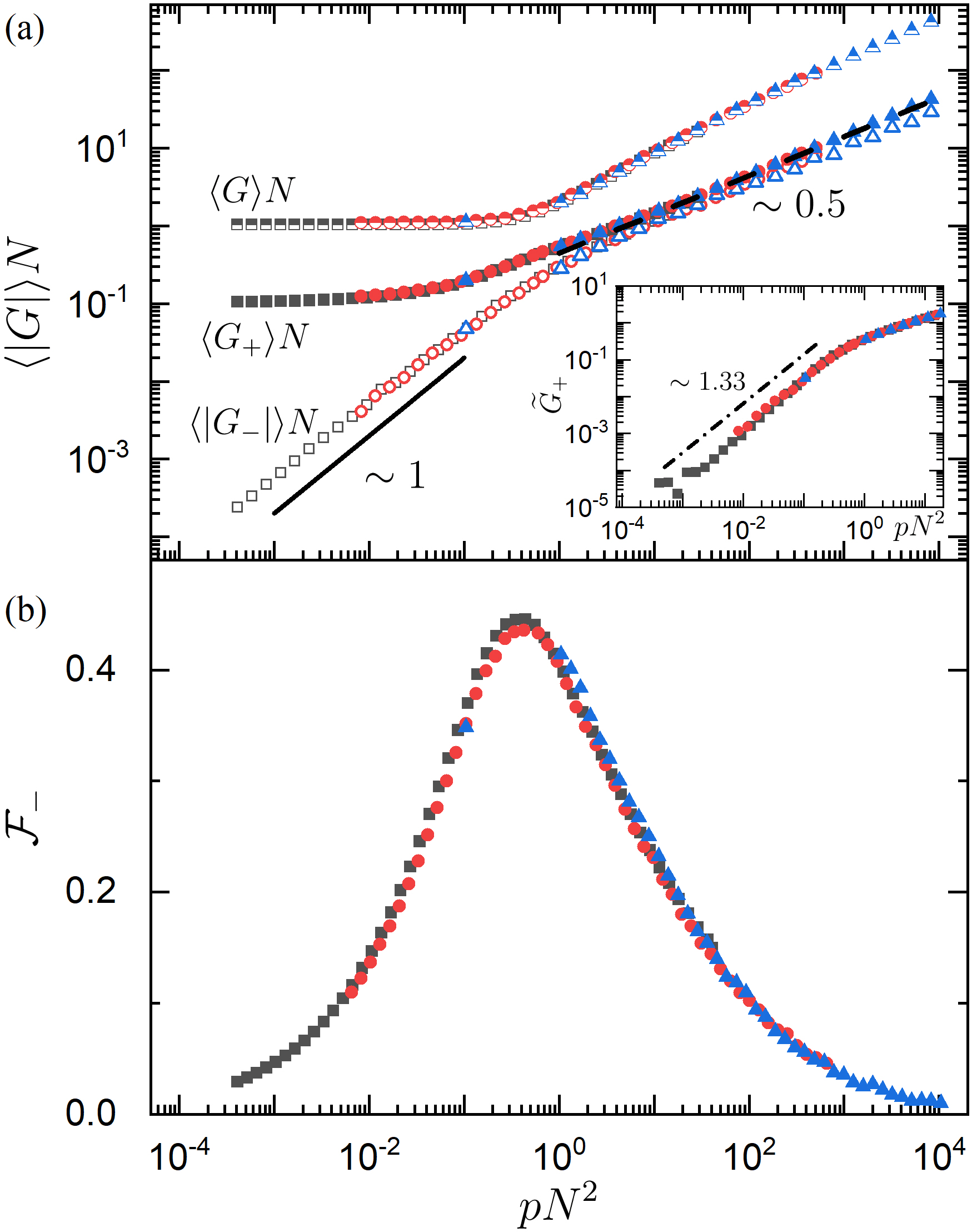}
\caption{(a) Ensemble-averaged positive ($\langle G_+\rangle$, solid symbols), negative ($\langle G_-\rangle$, open symbols), and total global shear moduli ($\langle G \rangle$, half-filled symbols) scaled by $N$ and plotted as a function of the scaled pressure $pN^2$ for $N=64$ (black squares), $256$ (red circles), and $1024$ (blue triangles). $\langle G \rangle N$ is multiplied by a factor of $10$ to improve visualization. The solid and dashed lines have slopes $1$ and $0.5$, respectively. The inset shows $\widetilde{G}_{+}=[\langle G_{+} \rangle-G_0/(1-{\cal F}_-)] N$ versus $pN^2$, where $G_0=\langle G_+\rangle$ and ${\cal F}_-=0$ in the $p=0$ limit.  The dashed line in the inset has slope $1.33$. (b) The fraction $\mathcal{F}_{-}$ of jammed packings with negative shear moduli  ($G<0$) plotted as a function of $pN^2$ for the same systems in (a).}
    \label{fig:scaling}
\end{figure}

Similar to Eq.~\ref{stress_tensor}, the virial stress tensor of each triangle $i$ is defined as
\begin{equation}
\Sigma^{\Delta}_{i\alpha\beta}=\frac{1}{2}
\sum_{m>n}  r_{mn\alpha} f_{mn\beta},
\label{eq:Stri}
\end{equation}
where $m$ and $n$ refer to the three disks forming a given Delaunay triangle $i$. Note that each contacting pair of disks is shared by two triangles and thus the stress from this contact contributes half to each triangle. The area factor in Eq.~\ref{stress_tensor} is not included in Eq.~\ref{eq:Stri} to simplify the classification of triangle types. (See Appendix~\ref{appxtristiff}.)

The virial stress and deformation gradient tensors for a subsystem $\ell$ that is composed of $n_\ell$ connected triangles are  
\begin{subequations}
\begin{align}
    {\overline \Sigma}^\ell&=\sum^{n_\ell}_i {\overline \Sigma}^{\Delta}_i,\label{eq:stresssub}\\
    {\overline F}^\ell&=\frac{1}{A^\ell}\sum^{n_\ell}_i A^{\Delta}_i {\overline F}^{^{\Delta}}_i,\label{eq:Fsub}
\end{align}
\end{subequations}
where ${\overline \Sigma}^{\Delta}_i$, ${\overline F}^{^{\Delta}}_i$, and $A^{\Delta}_i$ are the virial stress tensor, deformation gradient tensor, and area of triangle $i$, respectively, and $A^\ell=\sum^{n_\ell}_i A^{\Delta}_i$. 
We can substitute Eqs.~\ref{eq:stresssub} and~\ref{eq:Fsub} into Eqs.~\ref{eq:Etri},~\ref{eq:Smtltri} and~\ref{eq:stifftri} to obtain the Green-Lagrangian strain tensor ${\overline E}^\ell$, material stress tensor ${\overline \Sigma}^{m,\ell}$, and the associated stiffness tensor ${\hat C}^\ell$ of subsystem $\ell$. In Sec.~\ref{sectri}, we will consider a range of subsystems with different sizes, e.g. single Delaunay triangles, pairs of triangles that share one edge, polygons whose vertices correspond to a disk and its Voronoi-neighbor disks, and subsystems containing an average number of disks $n_{\rm sub} = 2 N/n^2$ whose centroids are located within squares of side length $L/n$. 

\begin{figure}[t]
\centering
    \includegraphics[width=0.425\textwidth]{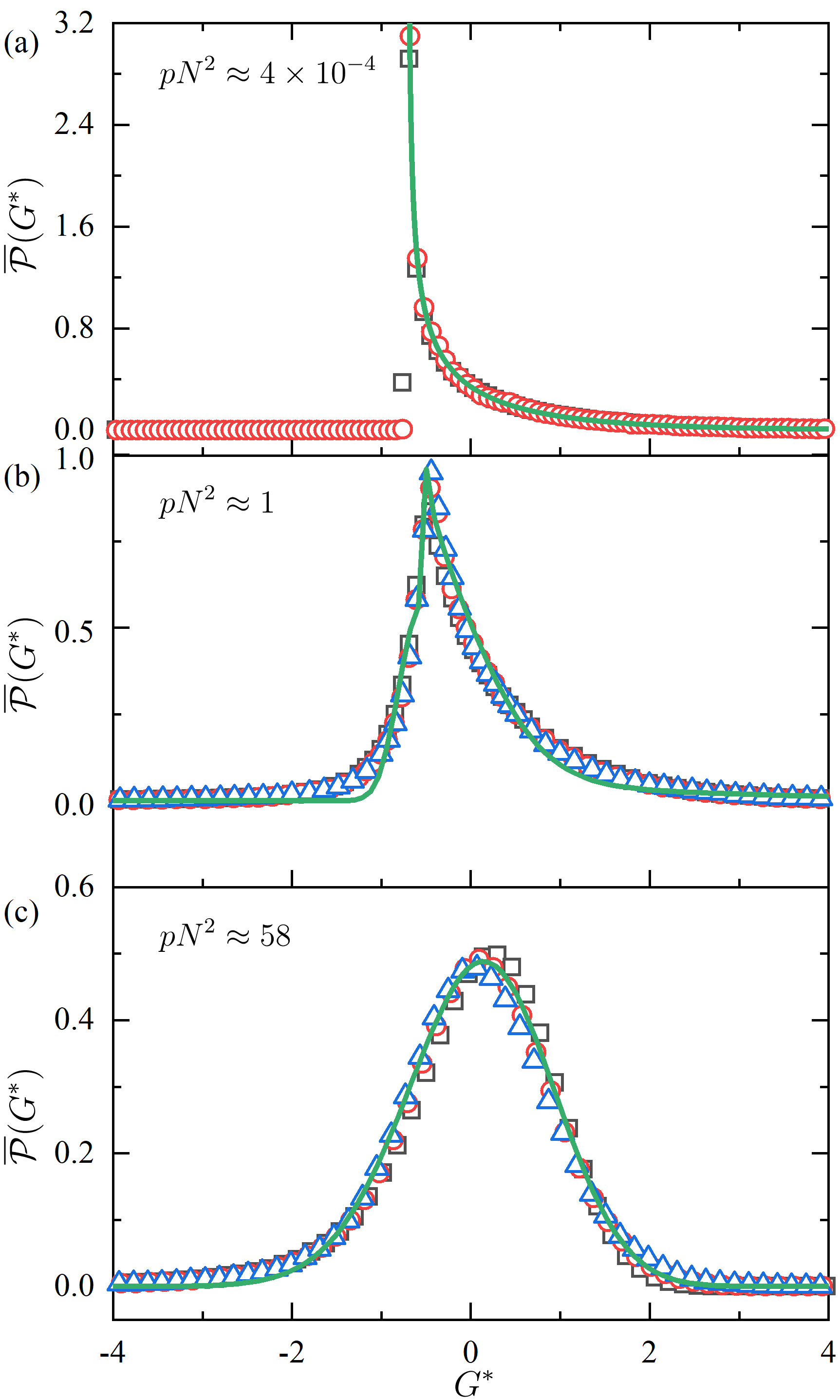}
\caption{The probability distribution ${\overline {\cal P}}(G^*)$ of shifted and normalized global shear moduli, where $G^*=(G-\langle G\rangle)/{\cal S}_{G}$ and ${\cal S}_G$ is the standard deviation in $G$ for jammed disk packings with (a) $pN^2=4\times 10^{-4}$, (b) $1$, and (c) $58$ and system sizes $N=64$ (black squares), $256$ (red circles), and $1024$ (blue triangles). The solid lines in (a) and (c) represent a Gamma distribution with shape parameter $k = 0.5$ (Eq.~\ref{eq:pdfG*}) and a skew-normal distribution (Eq.~\ref{eq:sn}), respectively. An interpolation between these two forms (Eq.~\ref{eq:lerp}) is shown as the solid line in (b). The parameters that specify the distributions in (b) and (c) are given in Table~\ref{tab:table1}.}
    \label{fig:pdfGpN2}
\end{figure}

\section{Results}
\label{results}

Our results are organized into three subsections. In Sec.~\ref{secglobal}, we describe how the inclusion of jammed packings with negative shear moduli affects the pressure dependence of the ensemble-averaged global shear modulus $\langle G \rangle$. We also show that the global shear modulus distribution ${\cal P}(G)$ collapses with $pN^2$ and its form varies from a right-skewed Gamma distribution in the $pN^2 \rightarrow 0$ limit to a left-skewed Gaussian distribution in the $pN^2 \gg 1$ limit. In Sec.~\ref{secaffinegam}, we describe the results for the distribution of the local shear moduli ${\cal P}(g^\ell)$ using the affine-strain method for decomposing the stress and strain tensors for each subsystem.  We show that the affine local shear moduli possess long-range spatial correlations over the full range of $pn_{\rm sub}^2$, where $n_{\rm sub}$ is the size of each subsystem. In Sec.~\ref{sectri}, we show that the form of ${\cal P}(g^\ell)$ differs for $g^\ell$ defined using the affine and non-affine methods.  The spatial correlations of $g^\ell$ defined using the non-affine method with Delaunay triangulation are much weaker than those defined using the affine method over the full range of $pn_{\rm sub}^2$.  We also show that jammed disk packings with global shear moduli $G>0$ can possess negative local shear moduli.

\begin{figure}[t]
    \centering
    \includegraphics[width=0.425\textwidth]{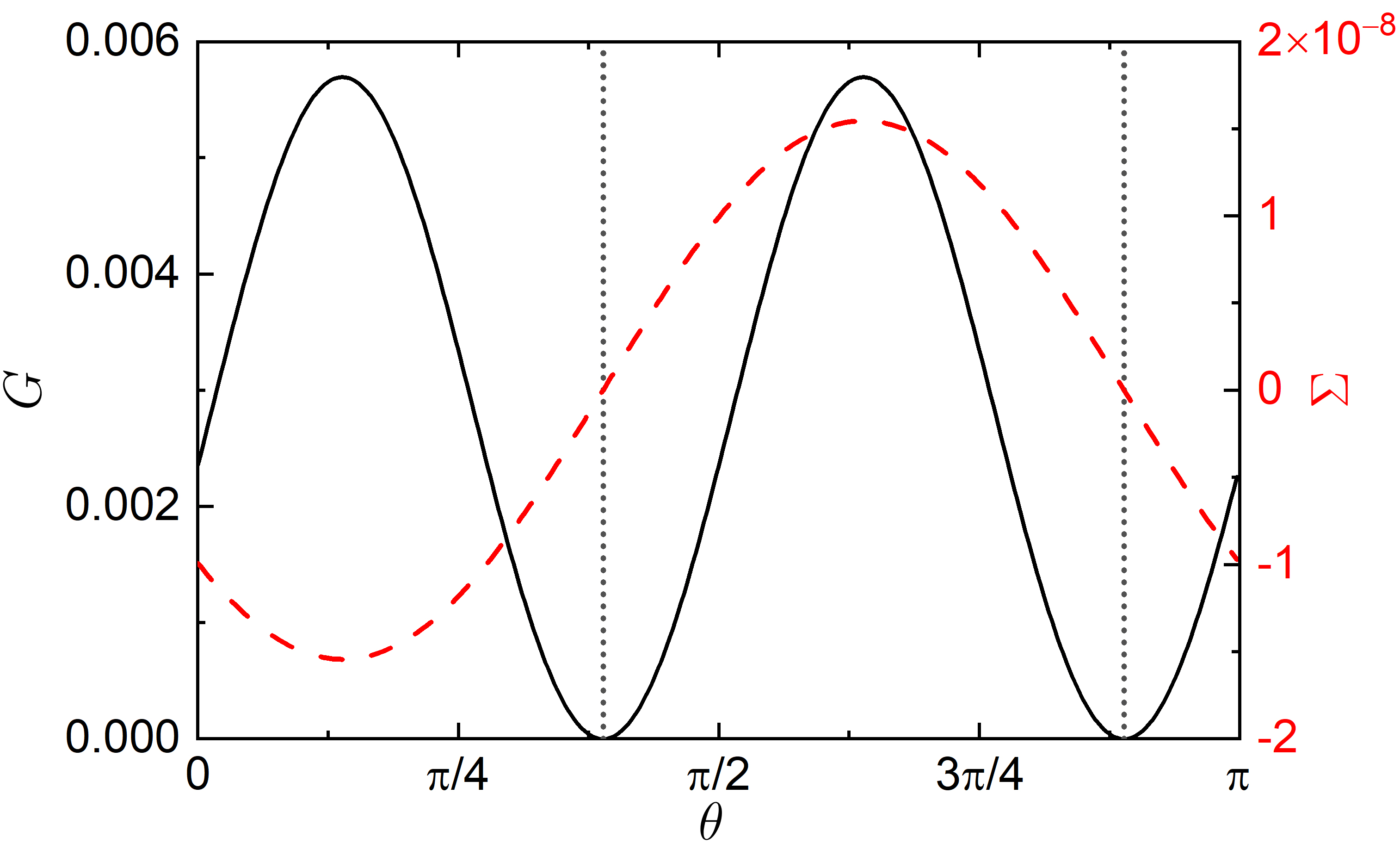}
\caption{The global shear modulus $G$ (solid line) and shear stress $\Sigma$ (dashed line) plotted as a function of the shear angle $\theta$ for a single disk packing in the $pN^2 \ll 1$ limit. The vertical dotted lines indicate values of the shear angle $\theta_c$ at which $\Sigma(\theta_c)=0$. At $\theta_c$, $G(\theta_c)$ is a minimum, which indicates that $\theta_G-\theta_S = \pi/8$.}
    \label{fig:offset}
\end{figure}

\subsection{Global shear modulus}
\label{secglobal}

In this section, we describe the pressure and system-size dependence of the global shear modulus probability distribution ${\cal P}(G)$ and the ensemble-averaged value, 
\begin{equation}
\label{decompose}
\langle G\rangle = (1-{\cal F}_-) \langle G_+ \rangle + {\cal F}_{-} \langle G_{-}\rangle,
\end{equation}
where ${\cal F}_-$ is the fraction of jammed packings with negative shear moduli, and $\langle G_+\rangle$ and $\langle G_-\rangle$ are the ensemble-averaged values of the positive and negative global shear moduli, respectively.  First, in Fig.~\ref{fig:scaling} (a), we show that $\langle G\rangle$ (as well as $\langle G_+\rangle$ and $\langle G_-\rangle$) collapse when plotted versus $pN^2$ as found previously~\cite{goodrich2012,goodrich2014}.  Previous computational studies of jammed sphere packings (with repulsive linear spring interactions) have also emphasized that the ensemble-averaged global shear modulus displays power-law scaling with pressure, $\langle G \rangle N \sim (pN^2)^{\beta}$, where $\beta \sim 0.5$, in the large-$pN^2$ limit~\cite{ohern2003,goodrich2012}.  However, in Fig.~\ref{fig:scaling} (a), the scaling exponent $\beta \gtrsim 0.5$ in the range $10 \lesssim pN^2 \lesssim 10^4$~\cite{wang2021shear}.  According to Eq.~\ref{decompose},
the scaling exponent $\beta$ can be larger than $0.5$ if ${\cal F}_-$ depends strongly on pressure, even when {\it both} $\langle G_+\rangle$ and $\langle G_- \rangle$ scale as $(p N^2)^{0.5}$ at large values of $pN^2$ ({\it cf.} Fig.~\ref{fig:scaling}  (a)). In particular, we show in Fig.~\ref{fig:scaling} (b) that the fraction ${\cal F}_{-}$ of packings with negative global shear moduli has strong $pN^2$ dependence; it forms a peak with ${\cal F}_- \sim 50\%$ for $pN^2 \sim 1$ and falls to zero for both smaller and larger values of $pN^2$.  Indeed, previous studies have shown that $\beta \approx 0.5$ for ensembles of jammed packings that are generated using the shear-stabilizing algorithm~\cite{dagoisbohy2012,goodrich2014}, which ensures that the jammed packings possess zero residual stress and $G>0$ in all directions.  

\begin{figure}[t]
    \centering
    \includegraphics[width=0.425\textwidth]{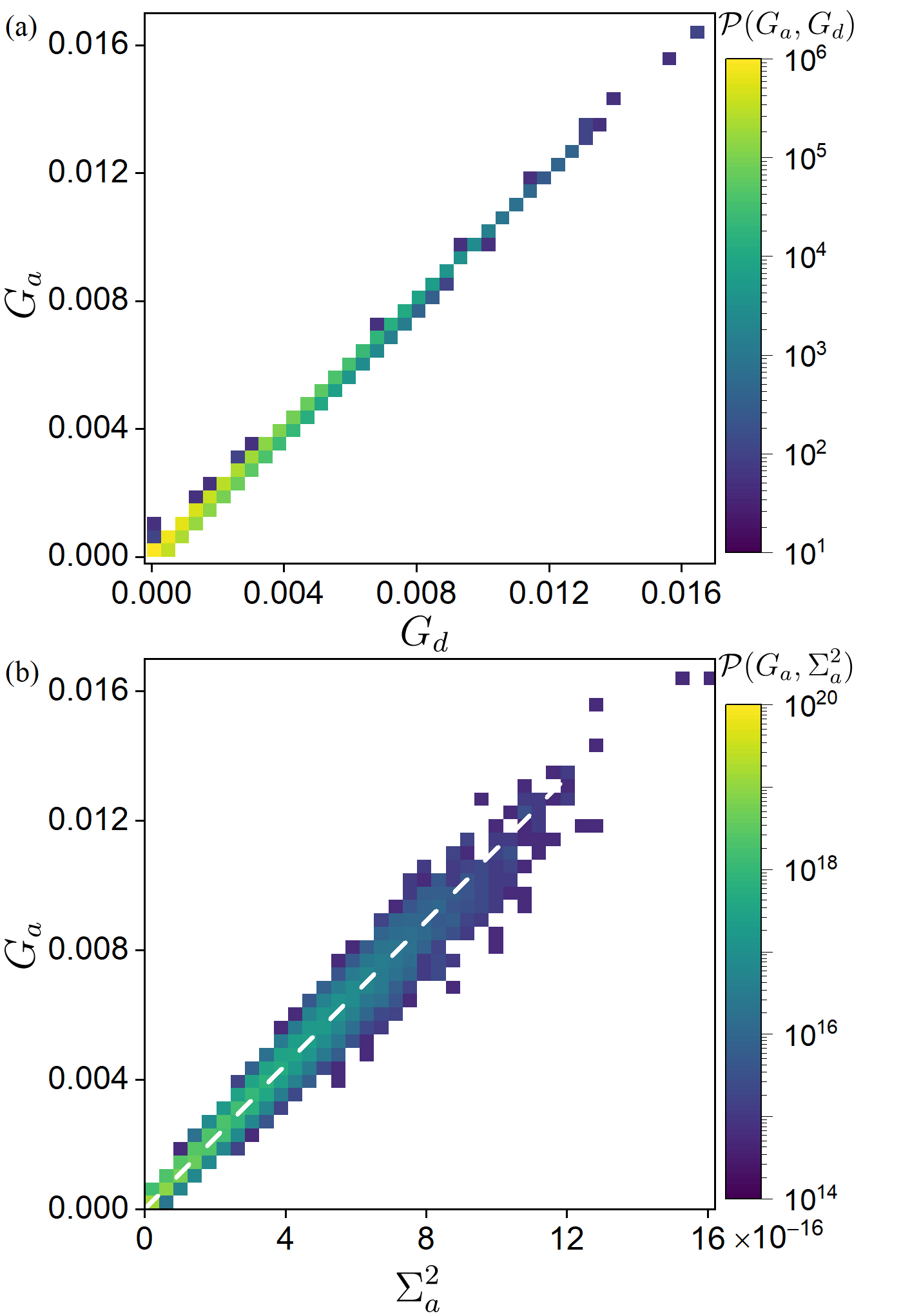}  
\caption{(a) Probability distribution ${\cal P}(G_a,G_d)$ for jammed disk packings with $N=64$ at low pressure $p=10^{-7}$. (b) Probability distribution ${\cal P}(G_a,\Sigma_{a}^2)$ for jammed packings with $N=64$ and $p=10^{-7}$. The dashed line obeys $G_a = A_c \Sigma_a^2$. In both panels, the probability increases from violet to yellow. }
    \label{fig:GaGdSa}
\end{figure}

In Fig.~\ref{fig:scaling} (a), we show that in the low-pressure limit the ensemble-averaged global shear modulus $\langle G\rangle$ tends to a constant $G_0 \sim N^{-1}$ that decreases to zero in the large-system limit~\cite{goodrich2012}. Previous studies of jammed packings of frictionless, spherical particles have shown that $(\langle G\rangle - G_0)N \sim (pN^2)^{\eta}$ with $\eta \sim 1$ for $pN^2 < 1$. In the current studies, we find that $[\langle G_{+} \rangle-G_0/(1-{\cal F}_-)] N \sim (pN^2)^{\eta_+}$, where $\eta_+ \sim 1.33$, and $\langle|G_-|\rangle N \sim pN^2$ in the low-pressure limit. Thus, the difference in the scaling exponents $\eta < \eta_+$ is caused by the occurrence of packings with negative shear moduli and we expect $\eta \sim \eta_+$ in the $pN^2 \ll 1$ limit, where ${\cal F}_- =0$.    

\begin{figure}[t]
    \centering
    \includegraphics[width=0.425\textwidth]{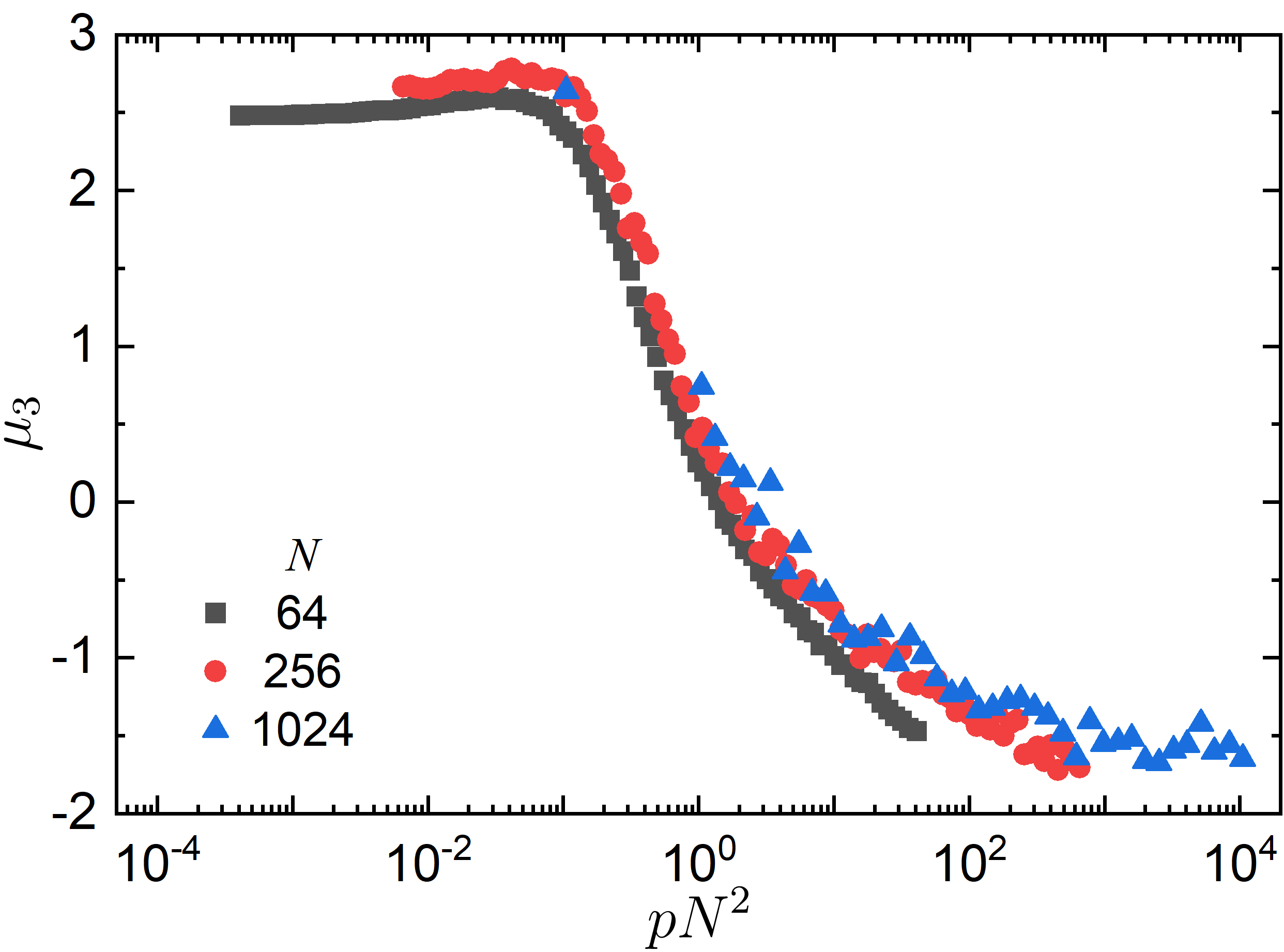}
\caption{The skewness ${\mu}_3$ (Eq.~\ref{skewness}) of the shifted and normalized distribution of global shear moduli ${\cal P}(G^*)$ plotted versus $pN^2$ for $N=64$, $256$, and $1024$.}
    \label{fig:skew}
\end{figure}

We have shown that the power-law scaling of the ensemble-averaged shear modulus depends on the fraction of jammed disk packings with negative shear moduli. We will now study the probability distribution of global shear moduli ${\cal P}(G)$ as a function of pressure and system size to determine the prevalence of $G<0$.  In Fig.~\ref{fig:pdfGpN2}, we show the shifted and normalized distributions ${\overline {\cal P}}(G^*)={\cal P}(G^*){\cal S}_G$,
where
\begin{equation}
\label{shifted}
    G^*=\frac{G-\langle G\rangle}{ {\cal S}_G},
\end{equation}
and ${\cal S}_G$ is the standard deviation of ${\cal P}(G)$.

As we found for the average values, the probability distribution ${\overline {\cal P}}(G^*)$ collapses at fixed $pN^2$ (at different values of $p$ and $N$). In the $pN^2 \gg 1$ limit (e.g. $pN^2 = 58$ in Fig.~\ref{fig:pdfGpN2} (c)), ${\overline {\cal P}}(G^*)$ obeys a skew-normal distribution (Eq.~\ref{eq:sn} in Appendix~\ref{appxdistG}) with negative skewness. See Table~\ref{tab:table1} for the specific parameters of the skew-normal distribution that describe ${\cal P}(G^*)$ in Fig.~\ref{fig:pdfGpN2} (c). In contrast, in the $pN^2 \ll 1$ limit, ${\overline {\cal P}}(G^*)$ obeys a Gamma distribution with shape parameter $k=0.5$ for $G^* > -\langle G\rangle/{\cal S}_G$ and is zero for $G^* < -\langle G\rangle/{\cal S}_G$, as shown in Fig.~\ref{fig:pdfGpN2} (a) for $pN^2 = 4 \times 10^{-4}$. (See Eq.~\ref{rewrite} in Appendix~\ref{appxdistG}.)

We now derive an expression for the probability distribution ${\cal P}(G)$ for disk packings in the $pN^2 \ll 1$ limit. As shown in Eq.~\ref{eq:stressG}, both the global shear modulus $G$ and shear stress $\Sigma$ vary sinusoidally with the shear angle $\theta$ (defined in Fig.~\ref{fig:shear}), which implies that the relation between $G$ and $\Sigma$ is a Lissajous curve~\cite{fahy1952} with an angular frequency ratio of $2$.  Using Eqs.~\ref{eq:stress} and~\ref{eq:G}, we find that $G$ and $\Sigma$ are related via
\begin{equation}
\begin{aligned}
G(\theta)= &\left(\frac{2 [\Sigma(\theta)]^2}{\Sigma_a^2}-1\right)G_a\sin 4(\theta_G-\theta_S)+G_d\\
&-\frac{2 \Sigma(\theta)}{\Sigma_a}\sqrt{1-\frac{[\Sigma(\theta)] ^2}{\Sigma_a^2}}G_a \cos4(\theta_G-\theta_S).
\end{aligned}
\label{eq:GSgen}
\end{equation}
We show in Fig.~\ref{fig:offset} that at jamming onset the difference in the phase shift between $G(\theta)$ and $\Sigma(\theta)$ satisfies $\theta_G-\theta_S = \pi/8$ and in Fig.~\ref{fig:GaGdSa} (a) we show that $G_a = G_d$ at jamming onset.  Thus, in the $pN^2 \ll 1$ limit, Eq.~\ref{eq:GSgen} becomes
\begin{equation}
    G=2\frac{G_a}{\Sigma_a^2}\Sigma^2.
    \label{eq:GS}
\end{equation} 

\begin{figure}[t]
    \centering    \includegraphics[width=0.425\textwidth]{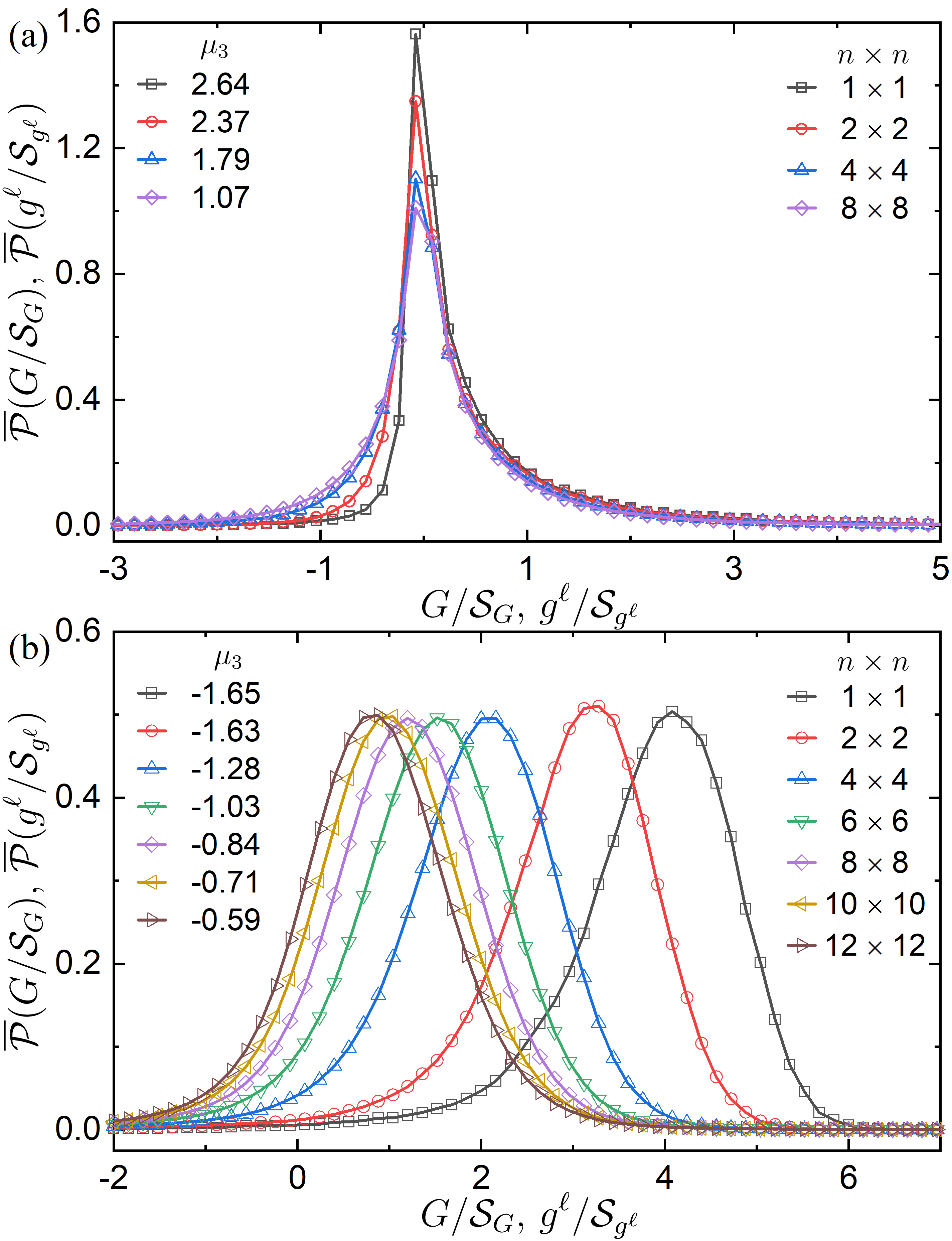}    
\caption{The probability distribution of normalized global shear moduli ${\overline {\cal P}}(G/{\cal S}_G)$ and probability distribution of normalized local shear moduli ${\overline {\cal P}}(g^\ell/{\cal S}_{g^\ell})$ obtained from the affine-strain method as a function of the $n\times n$ subsystem size and (a) $pN^2\approx 0.1$ and (b) $10^4$ for $N=1024$. The skewness ${\mu}_3$ of the distributions for each subsystem size is indicated.}
    \label{fig:pdflocalg}
\end{figure}
\begin{figure}[t]
    \centering
\includegraphics[width=0.425\textwidth]{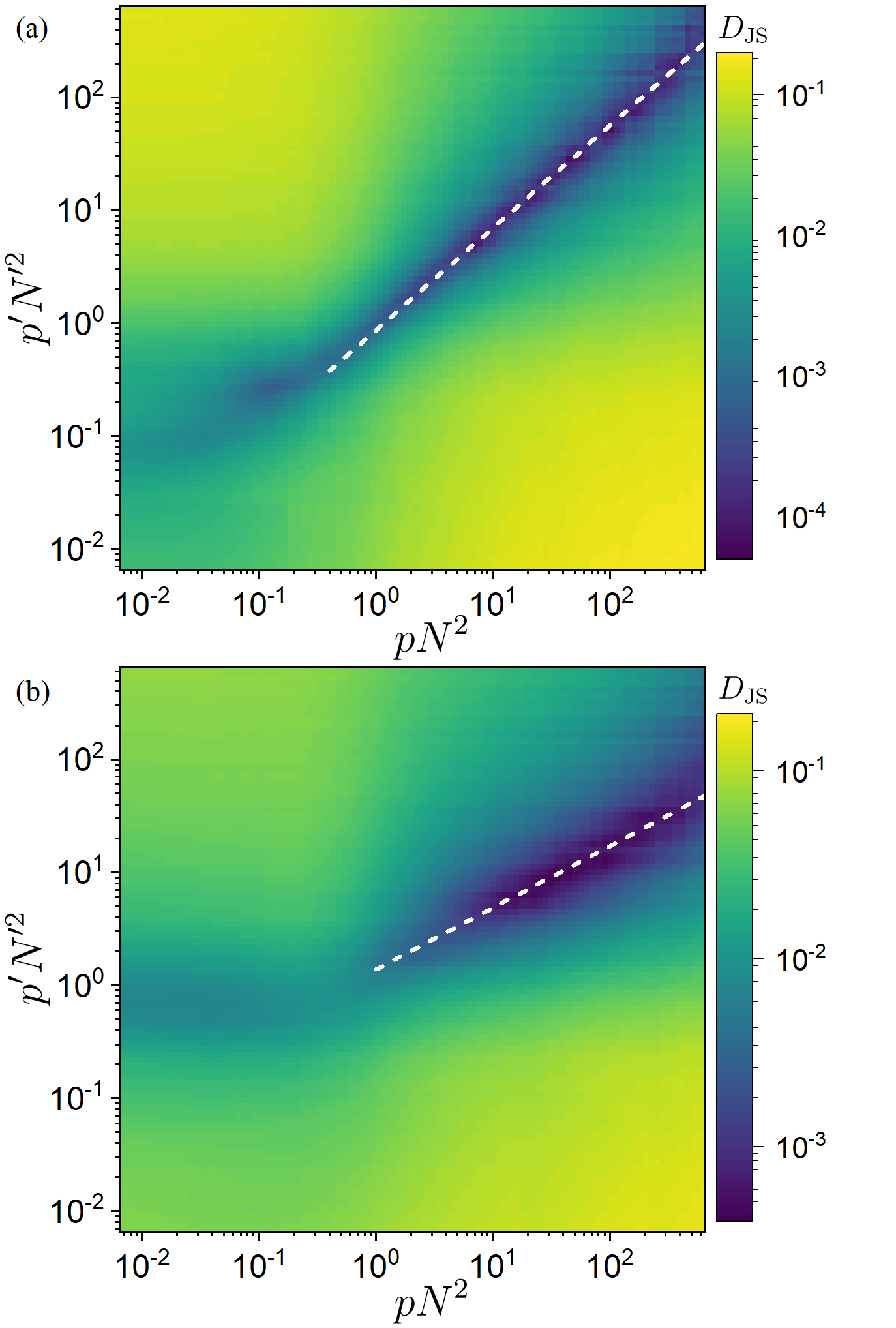}    
\caption{The Jensen-Shannon divergence $D_{JS}$ between the probability distribution of global shear moduli ${\overline {\cal P}}(G^*)$ at pressure $p'$ and system size $N'$ and the probability distribution of local shear moduli ${\overline {\cal P}}(g^{\ell*})$ (calculated using the affine-strain method) at pressure $p$ and system size $N$ for subsystem sizes (a) $n^2=4$ and (b) $25$.  $D_{JS}$ increases, i.e. the distributions become more dissimilar, from violet to yellow.  The dashed lines correspond to the power-law scaling $p^\prime N^{\prime2}\sim(pN^2)^\nu$, where $\nu \sim 0.91$ and $0.55$ in panels (a) and (b), respectively.}
    \label{fig:similar}
\end{figure}

Further, in Fig.~\ref{fig:GaGdSa} (b), we show that the amplitude of the shear modulus $G_a$ is proportional to $\Sigma_{a}^2$ in the $pN^2 \ll 1$ limit, $G_a = A_c \Sigma_a^2$, where $A_c \sim 1/p^2$ and $\Sigma_a^2 \sim p^2/N$. Also, previous studies have shown that the probability distribution of the shear stress for jammed disk packings generated by isotropic compression is Gaussian centered on $\Sigma=0$~\cite{sheng18}:
\begin{equation}
{\cal P}(\Sigma)=\frac{1}{\omega_s\sqrt{2\pi}}e^{-\frac{1}{2}(\Sigma/\omega_s)^2},
\label{eq:pdfSxy}
\end{equation}
where $\omega_s$ is the standard deviation. Using Eqs.~\ref{eq:GS} and~\ref{eq:pdfSxy}, we show in Appendix~\ref{appxPDF0} that the probability distribution of the global shear moduli is a Gamma distribution with shape parameter $k=0.5$ in the $pN^2 \ll 1$ limit: 
\begin{equation}
\begin{aligned}
{\cal P}_{\Gamma}(G)=\frac{1}{2 \omega_s\sqrt{\pi A_c G}}e^{-\frac{G}{4 A_c \omega_s^2}}.
\end{aligned}
\label{eq:pdfG}
\end{equation}
We can now rewrite Eq.~\ref{eq:pdfG} in terms of the shifted and normalized shear modulus $G^*$ in Eq.~\ref{shifted}: 
\begin{equation}
\begin{aligned}
{\overline {\cal P}}_{\Gamma}(G^*)=\frac{1}{\sqrt{\pi}\sqrt{1+\sqrt{2}G^*}}e^{-\frac{1}{2}\left(1+\sqrt{2}G^* \right)}. 
\end{aligned}
\label{eq:pdfG*}
\end{equation}
This expression is indicated by the solid line in Fig.~\ref{fig:pdfGpN2} (a). 

As shown in Fig.~\ref{fig:pdfGpN2}, ${\overline {\cal P}}(G^*)$ varies continuously with $pN^2$ from a Gamma distribution for $pN^2 \ll 1$ to a skew-normal distribution for $pN^2 \gg 1$. ${\overline {\cal P}}(G^*)$ at intermediate values of $pN^2$ can be approximated by a linear combination of ${\overline {\cal P}}_{\Gamma}(G^*-G_m^*)$ (where $G_m^*$ is the location of the maximum of ${\overline {\cal P}}(G^*)$ and ${\overline {\cal P}}_{SN}(G^*)$, as shown in Fig.~\ref{fig:pdfGpN2}(b). The best-fit parameters for ${\overline {\cal P}}(G^*)$ in Fig.~\ref{fig:pdfGpN2} are listed in Table~\ref{tab:table1}. In Fig.~\ref{fig:skew}, we show the skewness
\begin{equation}
\label{skewness}
{\mu}_3=\frac{\langle (G-\langle G\rangle)^3\rangle}{{\mathcal{S}_G}^{3}},
\end{equation}
of ${\overline {\cal P}}(G^*)$ as a function of $pN^2$. The skewness is positive in the $pN^2 \ll 1$ limit since ${\overline {\cal P}}(G^*)$ is a Gamma distribution, it slightly increases with $pN^2$ for $pN^2<0.1$, and then it decreases rapidly for $pN^2 > 0.1$.  The skewness becomes negative and reaches a plateau value $\mu_3 \sim -1.5$ in the $pN^2 \gg 1$ limit.

\begin{table}[b]
\caption{\label{tab:table1}
The parameters that determine the shape of the probability distributions of the global shear moduli $\overline{{\cal P}}(G^*)$ in Fig.~\ref{fig:pdfGpN2}, where $G^* = (G-\langle G\rangle)/{\cal S}_G$, ${\cal S}_G$ is the standard deviation of $G$, $\mu_{SN}$ is the shape parameter of the skew-normal distribution, $\overline{{\cal P}}_{SN}(G^*)$, and $0 \le s \le 1$ determines the relative contribution of $\overline{{\cal P}}_{\Gamma}(G^*-G_m^*)$ (where $G_m^*$ is the location of the peak in $\overline{{\cal P}}(G^*)$) and $\overline{{\cal P}}_{SN}(G^*)$ to $\overline{{\cal P}}(G^*)$ ({\it cf.} Eqs.~\ref{rewrite},~\ref{eq:sn}, and~\ref{eq:lerp}). ${\cal S}^{SN}_{G^*}$ and ${\cal S}^{\Gamma}_{G^*}$ are the standard deviations of the $\overline{{\cal P}}_{SN}(G^*-G_m)$ and $\overline{{\cal P}}_{\Gamma}(G^*)$ contributions to $\overline{{\cal P}}(G^*)$.  
}
\begin{tabular}{|c|cc|ccc|c|}
 \hline
 \multirow{2}{*}{Distribution} & \multicolumn{2}{c|}{$\overline{{\cal P}}_\Gamma(G^*-G_m^*)$}  & \multicolumn{3}{c|}{$\overline{{\cal P}}_{SN}(G^*)$} & \multirow{2}{*}{$s$}\\
                     & $G_m^*$ & ${\cal S}_{G^*}^{\Gamma}$ & $\langle G^* \rangle^{SN}$ & ${\cal S}_{G^*}^{SN}$ & $\mu_{SN}$ & \\
\hline
Fig.~\ref{fig:pdfGpN2} (b) & -0.571 & 2.446 & -0.184 & 0.503 & 4.314 & 0.665\\
Fig.~\ref{fig:pdfGpN2} (c) & - & - & 0.0712 & 0.735 & -1.032 & 0\\
 \hline
\end{tabular}
\end{table}

\begin{figure}[t]
    \centering
    \includegraphics[width=0.425\textwidth]{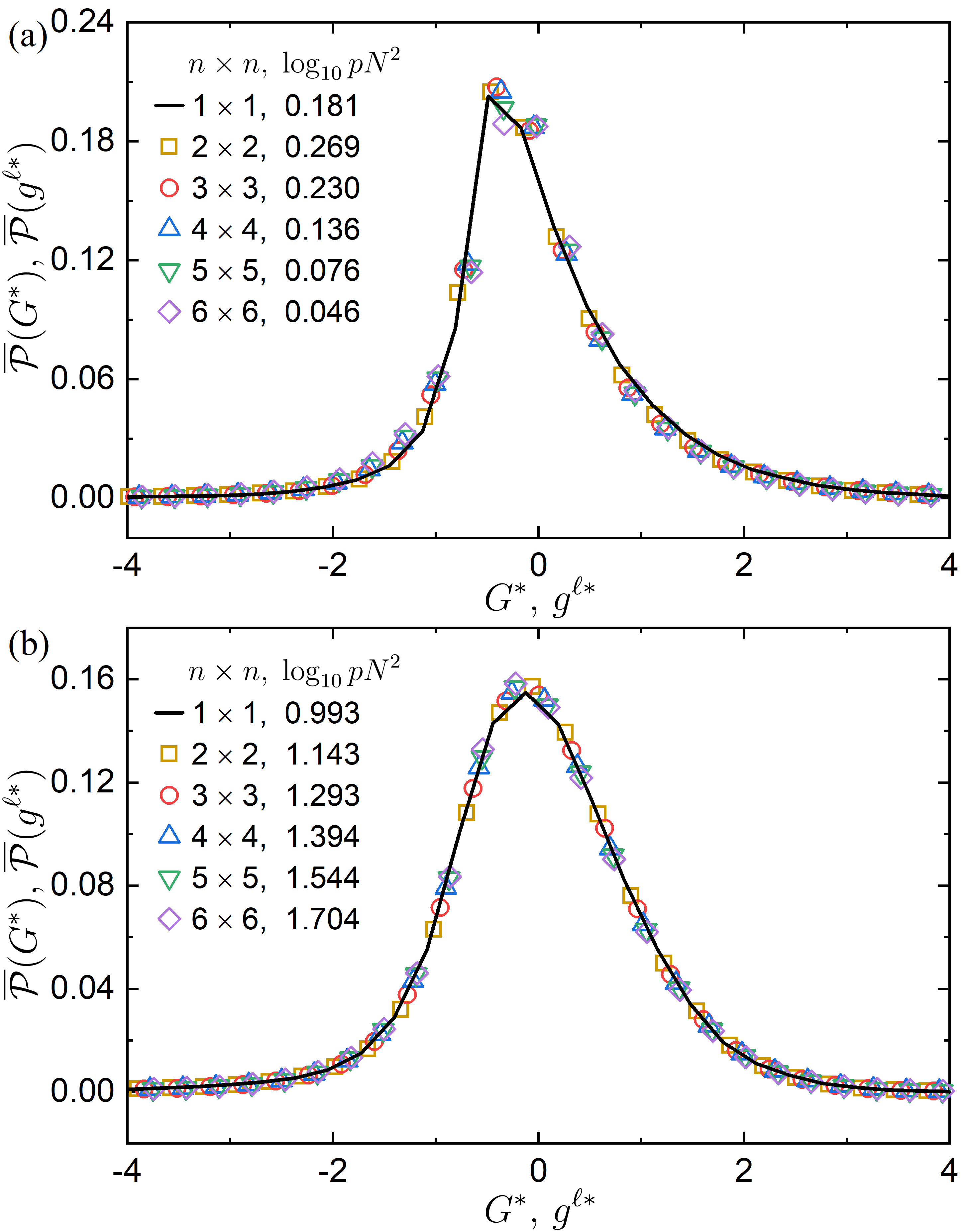}
\caption{Probability distributions of the normalized and shifted global ${\overline {\cal P}}(G^*)$ and local shear moduli ${\overline {\cal P}}(g^{\ell*})$ (calculated using the affine-strain method) for values of $p$ and $N$ and $p'$ and $N'$ that yield similar distributions.  The subsystem sizes $1 \le n^2 \le 36$ and values of $pN^2$ are indicated. We show $p'N^{'2} \approx 1.5$ and $10$ in panels (a) and (b), which determine the shape of ${\overline {\cal P}}(G^*)$.}
    \label{fig:simexamp}
\end{figure}

\subsection{Local shear moduli $g^\ell$ defined using the affine-strain method}
\label{secaffinegam}

In this section, we focus on the {\it local} shear moduli of jammed disk packings. In particular, we investigate whether the local shear moduli of jammed disk packings mimic the distribution of global shear moduli.  For example, do jammed disk packings possess negative local shear moduli? We first calculate the local shear moduli $g^\ell$ using the affine-strain method and determine the probability distribution ${\overline {\cal P}}(g^\ell)$ and spatial correlations in $g^\ell$ as a function of $p$ and $N$. 

\begin{figure}[t]
    \centering
    \includegraphics[width=0.425\textwidth]{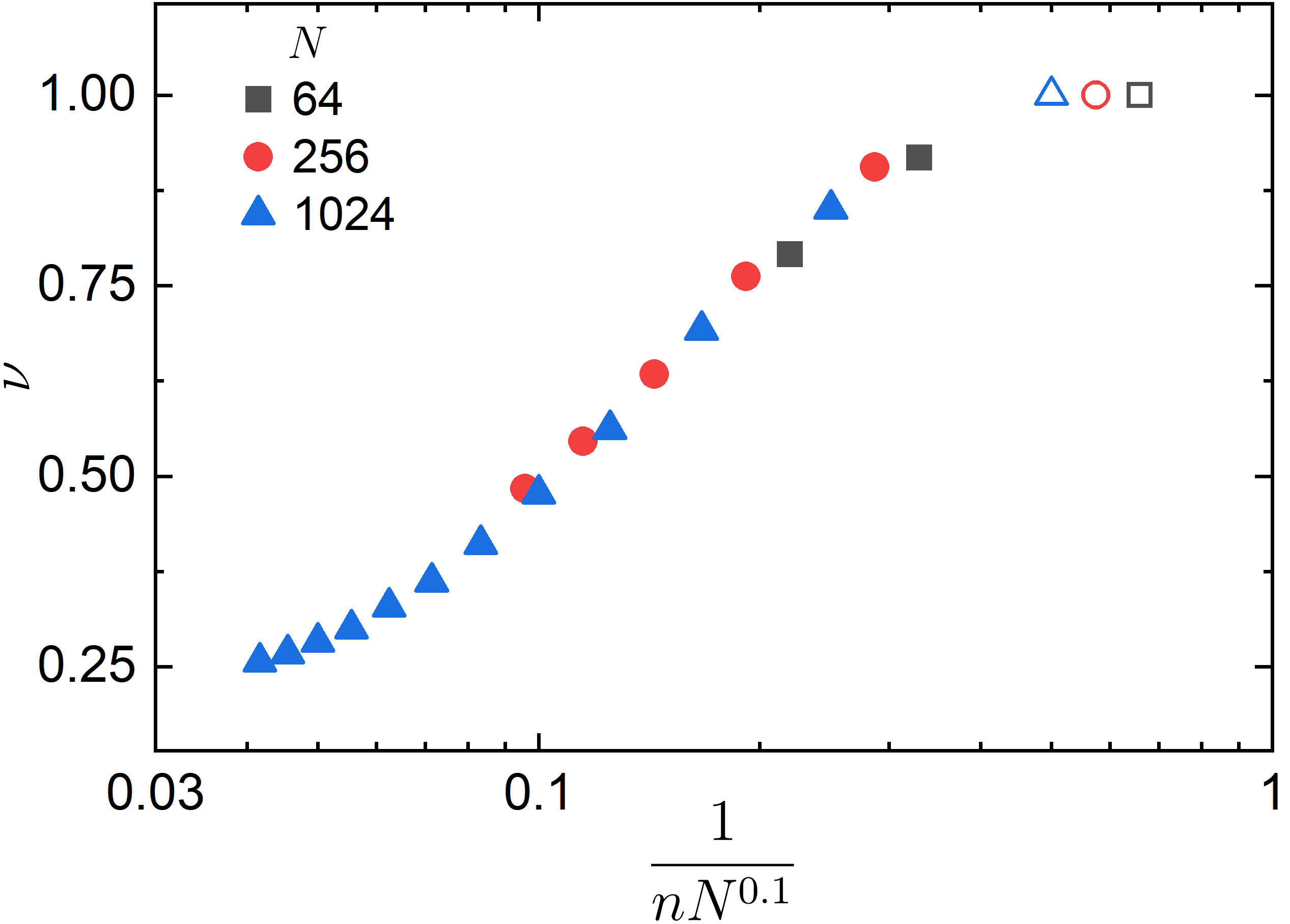}    
\caption{The power-law scaling exponent $\nu$ (filled symbols) in Eq.~\ref{scaling} that relates pairs of pressures and system sizes that yield matching distributions for the global and local shear moduli (i.e. $D_{JS} \lesssim 10^{-2}$) plotted as a function of $1/n$, where $n_{\rm sub}=N/n^2$ is the subsystem size.  $\nu$ is only weakly dependent on system size when we include the factor of $N^{-0.1}$. The open symbols with $\nu=1$ correspond to comparisons of the distributions of global shear moduli at different values of $p$ and $N$, but the same values of $pN^{2}$.}
    \label{fig:simexp}
\end{figure}

\subsubsection{Probability distribution of local shear moduli ${\overline {\cal P}}(g^\ell)$}

The affine-strain method for calculating the local shear moduli of a jammed disk packing assumes that each of the $n\times n$ subsystems experiences the same simple shear strain $\gamma$. In Sec.~\ref{localg}, we defined $g^\ell=d\Sigma^l/d\gamma$, where the local shear stress $\Sigma^l$ is given by Eq.~\ref{localstress}. The area-weighted sum of $g^\ell$ over all subsystems yields the global shear modulus $G$. 
In Fig.~\ref{fig:pdflocalg}, we show the probability distribution of local shear moduli ${\overline {\cal P}}(g^\ell)$ as a function of subsystem size $n_{\rm sub} = N/n^2$ at $pN^2 \approx 0.1$ and $\approx 10^4$. At small values of $pN^2$, the maximum in ${\overline {\cal P}}(g^\ell)$ remains roughly unchanged as a function of subsystem size. The skewness of ${\overline {\cal P}}(g^\ell)$ decreases with decreasing subsystem size due to an increasing fraction of negative local shear moduli, $g^\ell <0$.  Thus, jammed packings with $G>0$ in the $pN^2 \ll 1$ can contain local regions with negative local shear moduli. 
At large values of $pN^2$, the peak position shifts to smaller $g^\ell$ and ${\overline {\cal P}}(g^\ell)$ becomes more symmetric as the subsystem size decreases, as shown in Fig.~\ref{fig:pdflocalg} (b). For all values of $pN^2$, ${\overline {\cal P}}(g^\ell)$ is more symmetric than the distributions of the global shear moduli. This result raises the question of whether there is a combination of $p$, $N$, and $n_{\rm sub}$ at which the probability distributions of global and local shear moduli have the same form. 

To quantitatively compare two probability distributions $P_1(x)$ and $P_2(x)$, where $x = g^{\ell*}$ or $G^*$, we will calculate their Jensen–Shannon divergence~\cite{manning1999}, 
\begin{equation}
    D_{\rm JS}(P_1, P_2)=\frac{1}{2}\left(D_{\rm KL}(P_1,P_M)+D_{\rm KL}(P_2, P_M)\right),
\end{equation}
where $P_M=\frac{1}{2}(P_1+P_2)$,
\begin{equation}
    D_{\rm KL}(P_1,P_M) = \int P_1(x)\log_2\left(\frac{P_1(x)}{P_M(x)}\right)dx,
\end{equation}
and $D_{\rm JS}(P_1,P_2)$ is bounded between $0$ (when $P_1=P_2$) and $1$ (when there is no similarity between $P_1$ and $P_2$). 

\begin{figure}[t]
    \centering
    \includegraphics[width=0.425\textwidth]{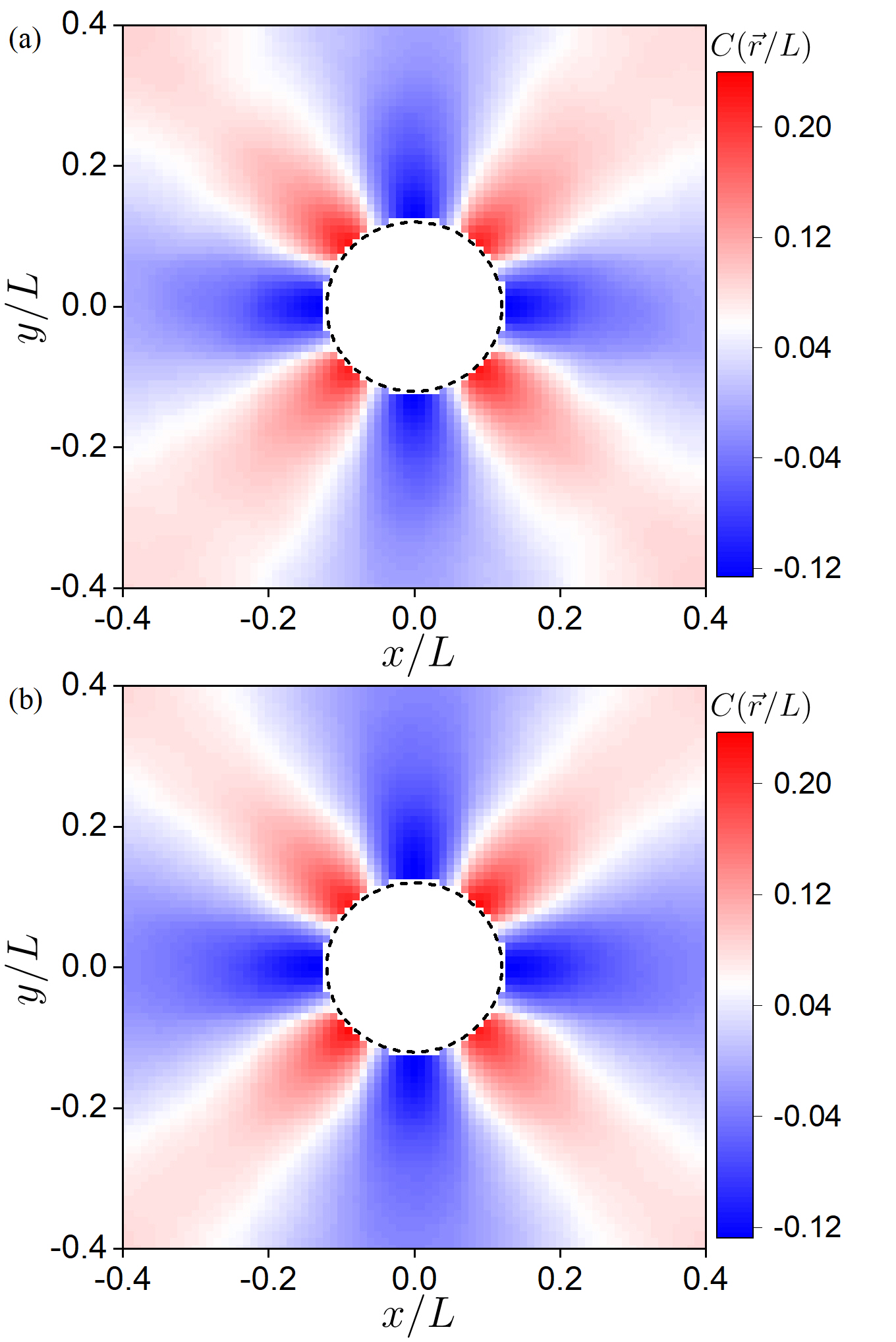}
\caption{Spatial correlation function $C({\vec r}/L)$ of the shifted and normalized local shear moduli $g^{\ell*}$ (calculated using the affine-strain method) for jammed disk packings with $n^2=144$ and pressures: (a) $pN^2=10^{-1}$ and (b) $10^{4}$. We do not display correlations for $r <\sqrt{2}L/n$ in the inner circular region.}
    \label{fig:corrg-affine}
\end{figure}

\begin{figure}[ht]
    \centering
    \includegraphics[width=0.4\textwidth]{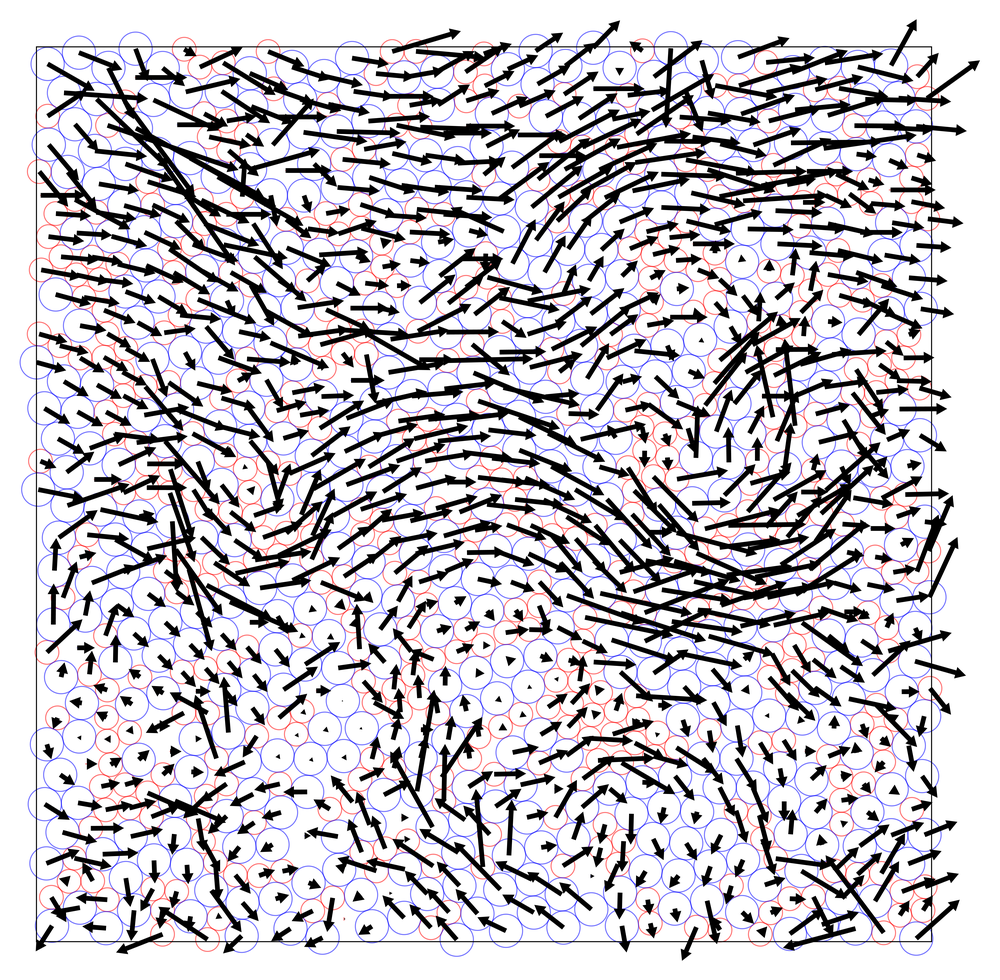}
\caption{The displacement field ${\vec r}_i(\gamma)-{\vec r}_i(0)$ for an $N=1024$ jammed disk packing at $pN^2 = 1$ after imposing a simple shear strain $\gamma = 10^{-9}$ followed by potential energy minimization.}
\label{fig:disfield}
\end{figure}

In Fig.~\ref{fig:similar}, we determine $D_{\rm JS}$ between ${\overline {\cal P}}(G^*)$ for jammed disk packings at pressure $p'$ and system size $N'$ and ${\overline {\cal P}}(g^{\ell*})$ for jammed disk packings at pressure $p$ and system size $N$ using subsystems with $n^2=4$ and $25$. 
For $pN^2>1$, one can identify values of $p'$ and $N'$ for which the distribution of global shear moduli ${\overline {\cal P}}(G^*)$ matches the distribution of local shear moduli ${\overline {\cal P}}(g^{\ell*})$ obtained from jammed disk packings at $p$ and $N$.  Examples of the matching pairs of distributions are shown in Fig.~\ref{fig:simexamp} for $p'N'^2 \approx 1.5$ and $10$. We find that the pairs $p$ and $N$ and $p'$ and $N'$ that yield similar distributions obey the following scaling relation: 
\begin{equation}
\label{scaling}
 p^\prime N^{\prime2}=A(pN^2)^\nu, 
\end{equation}
where $A$ is nearly constant over the range of subsystem and system sizes studied. In Fig.~\ref{fig:simexp}, we show that the power-law scaling exponent $\nu$ increases with increasing subsystem size with a weak overall system-size dependent correction.  Note that the range of $pN^2$ values over which $D_{\rm JS} \lesssim 10^{-2}$ decreases with increasing $n^2$. In particular, for $pN^2<1$, it is difficult to identify pairs of $p'$ and $N'$ and $p$ and $N$ at which the distributions of local and global shear moduli are similar.  The distributions of the local and global shear moduli become different in the $pN^2 \ll 1$ limit because $G>0$ for all jammed disk packings in that limit, yet as the subsystems become smaller, it is more likely for $g^\ell < 0$.

\subsubsection{Spatial correlations of local shear moduli $g^\ell$}

In Fig.~\ref{fig:corrg-affine}, we show the spatial correlation function of the shifted and normalized local shear moduli $C({\vec r})=\langle g^{\ell*}(0)g^{\ell*}(\vec{r})\rangle$ (using the affine-strain method) for subsystems with $n^2=144$ and pressures $pN^2 =10^{-2}$ and $10^4$. Over the full range of $pN^2$, we find that $C({\vec r})$ displays long-range four-fold spatial correlations. Previous studies have also found long-ranged spatial correlations in the local shear stress in zero-temperature amorphous solids~\cite{lemaitre2014Sigma_corr}. The long-range, angle-dependent spatial correlations imply that the size of the correlations will depend on the shape of the subsystems that are used to calculate the local shear modulus. For example, we have found that $C({\vec r})$ is significantly different for jammed packings decomposed into $n^2$ square subsystems with side lengths $L/n$ and into $n^2$ rectangular subsystems with side lengths $L/n^2$ and $L$. 

\subsection{Local shear moduli $g^\ell$ defined using the Delaunay triangulation method}
\label{sectri}

In the previous section, we focused on local shear moduli calculated using the affine-strain method. However, the disks in jammed packings have significant nonaffine motion in response to applied simple shear deformations~\cite{maloney,zaccone2011,richard20,jin2021}, as shown in Fig.~\ref{fig:disfield}. In this section, we characterize the local shear moduli of jammed disk packings using the Delaunay triangulation method to accurately define the local strain in each subsystem. We calculate the distribution of local shear moduli as a function of the size and shape of the subsystem, including triangles, polygons, and squares.  In addition, we determine the spatial correlations of the local shear moduli as a function of $pN^2$. 

\subsubsection{Types of Delaunay triangles}

We first consider the local shear moduli of subsystems composed of single triangles obtained from Delaunay triangulation of the disk centers in jammed disk packings. There are several types of triangles that can be obtained from Delaunay triangulation of binary disk packings, and we will classify them based on the form of the triangle stiffness matrix (Eq.~\ref{eq:C}). First, we do not consider triangles formed from three disks with no mutual contacts since they would have zero local stress. We define triangle type-1 as triangles with a single contact among the three disks. This triangle type can include all possible disk size combinations since in this case the stiffness matrix is the same for triangles with three large disks, three small disks, two small disks and one large disk, and two large disks and one small disk. We define triangle type-2 as triangles with two contacts among any of the three disks since the stiffness matrix again does not depend on the size combinations. For triangle types-3, -4, and -5, all disks are in contact with each other.  For type-3, all three disks are the same size. For type-4, two of the disks are small and one disk is large. For type-5, two of the disks are large and one is small. The triangle types are displayed in Appendix~\ref{appxtristiff}.

The stiffness matrix depends on each triangle's orientation. Thus, in Appendix~\ref{appxtristiff}, we first calculate the reference stiffness tensor ${\hat C}_{i0}^\Delta$ for each of the five triangle types in a specific reference orientation. We then calculate the stiffness matrix for each triangle $i$ in jammed disk packings using Eq.~\ref{eq:stifftri} and transform ${\hat C}_i^\Delta$ to the reference orientation using ${\hat C}^{\Delta}_{i,{\cal R}} = \mathcal{R}{\hat C}_i^\Delta \mathcal{R}^{\rm T}$, where 
\begin{equation}
\mathcal{R}=
\begin{bmatrix}
 \cos ^2\alpha_r & \sin ^2\alpha_r & \sin 2\alpha_r \\
 \sin ^2\alpha_r & \cos ^2\alpha_r & -\sin 2\alpha_r \\
 -\frac{1}{2}\sin 2\alpha_r & \frac{1}{2}\sin 2\alpha_r & \cos 2\alpha_r \\
\end{bmatrix}
\label{alpha}
\end{equation}
and $\alpha_r$ is the rotation angle that takes triangle $i$ from the orientation in the jammed disk packing to the reference orientation in Appendix~\ref{appxtristiff}.

\begin{figure}[t]
    \centering
    \includegraphics[width=0.425\textwidth]{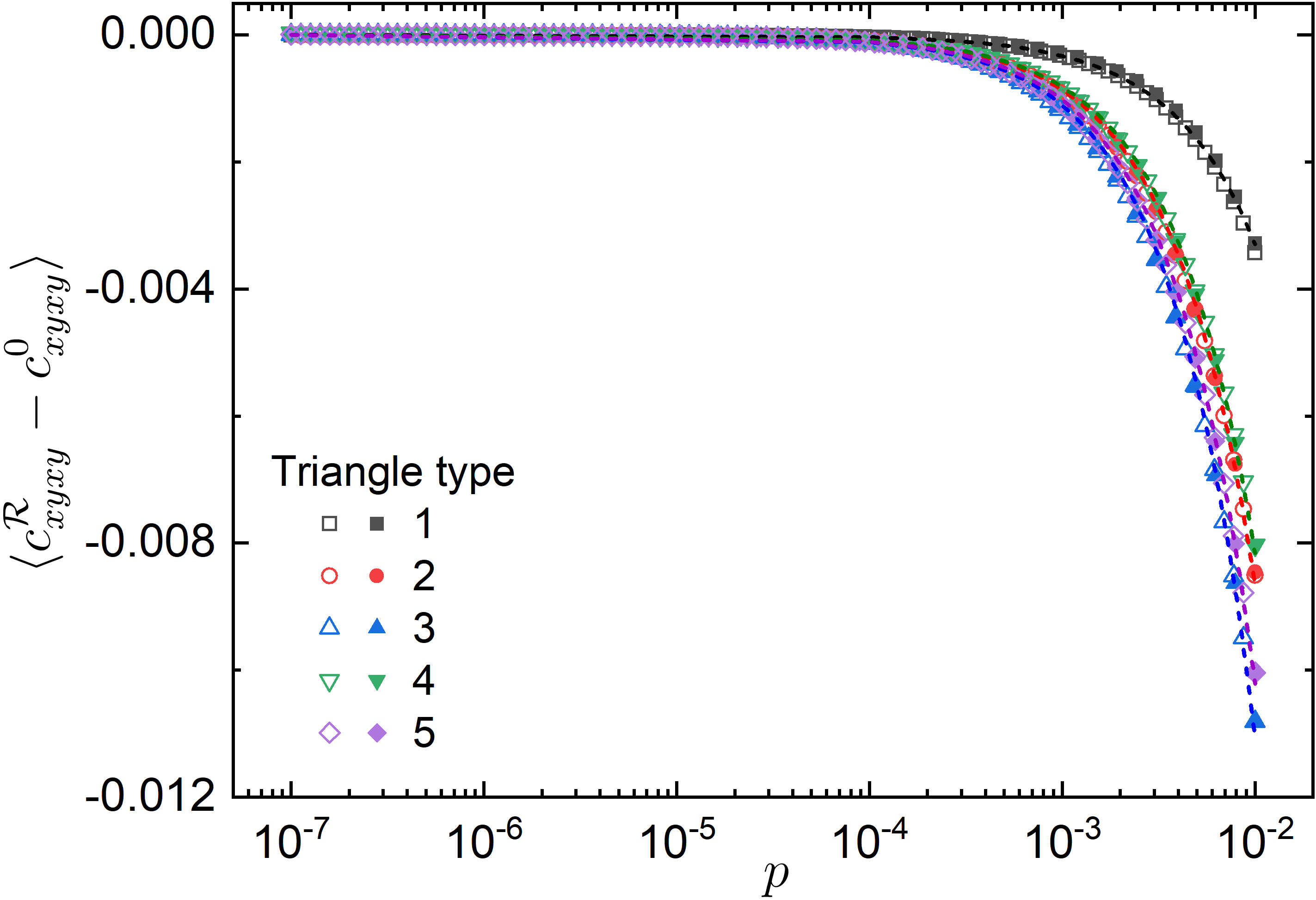}    
\caption{The average difference $\langle c_{xyxy}^{\cal R}- c_{xyxy}^0\rangle$ in the stiffness matrix components between single Delaunay triangles in jammed disk packings and the corresponding reference triangles plotted as a function of pressure $p$ for each triangle type in Appendix~\ref{appxtristiff}. Open and filled symbols indicate $N=256$ and $1024$, respectively, and the dashed lines indicate best fits to $\langle c_{xyxy}^{\cal R}- c_{xyxy}^0 \rangle= -\lambda p$.}
    \label{fig:c33diff}
\end{figure}

\begin{figure}[ht]
    \centering
    \includegraphics[width=0.425\textwidth]{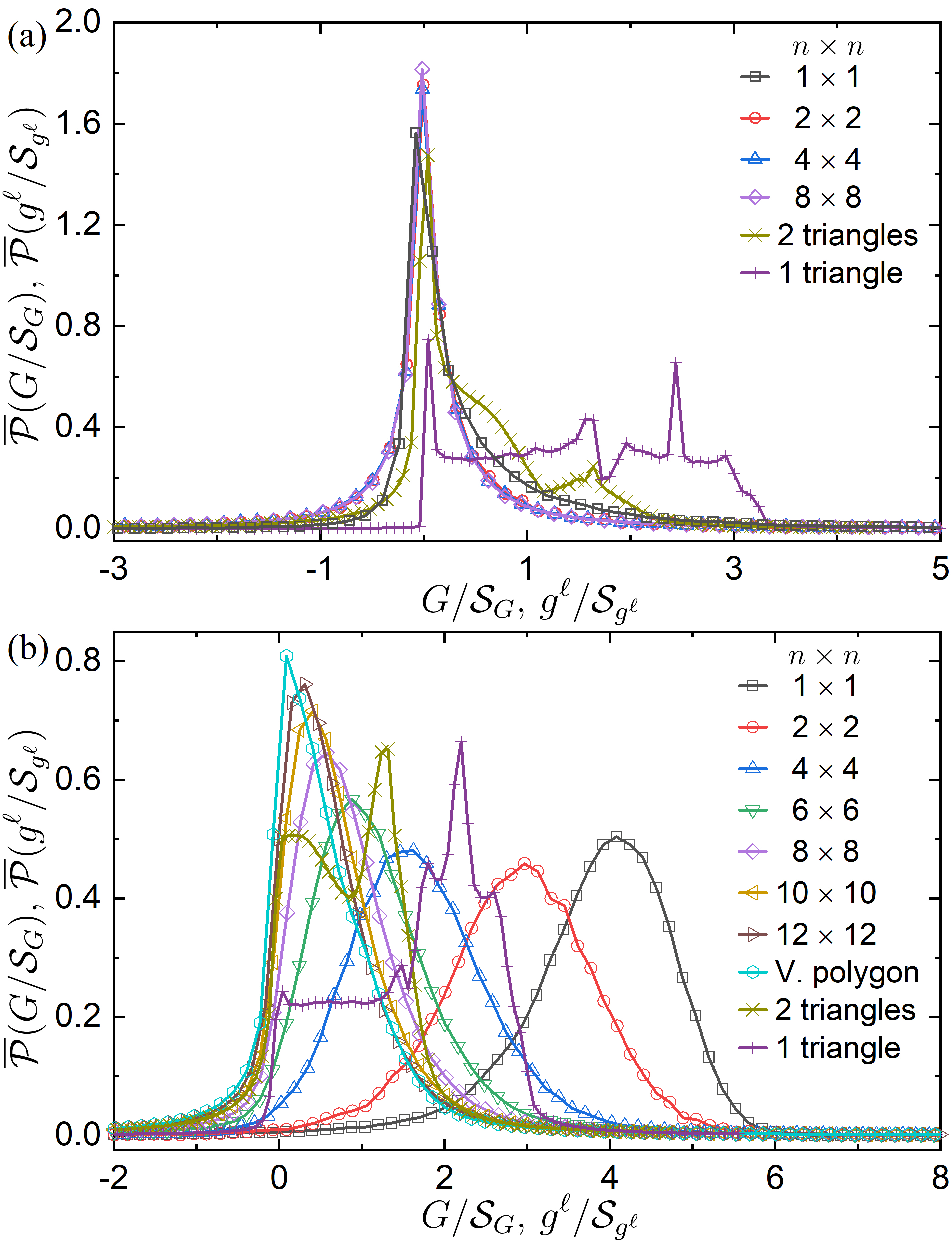} 
\caption{Probability distributions of the normalized global ${\overline {\cal P}}(G/{\cal S}_G)$ and local shear moduli ${\overline {\cal P}}(g^\ell/{\cal S}_{g^\ell})$ obtained via Delaunay triangulation for different subsystem shapes and sizes, including single triangles, two adjacent triangles, Voronoi polygons, and squares with $n^2 = 1$, $4$, $16$, $36$, $64$, $100$, and $144$ at (a) $pN^2\approx 0.1$ and (b) $10^4$ and $N=1024$.}
    \label{fig:pdfgtri}
\end{figure}

In Fig.~\ref{fig:c33diff}, we show the ensemble-averaged $xyxy$-component of the difference in the stiffness tensors, ${\hat C}^{\Delta}_{i,{\cal R}}-{\hat C}^{\Delta}_{i,0}$, for triangle $i$ in a given jammed packing and the corresponding reference triangle as a function of pressure. (Note that when a triangle changes type due to a particle rearrangement during compression, we stop measurements on that particular triangle.) Similar to the pressure dependence of the global shear modulus within geometrical families~\cite{xu2006measurements,vanderwerf2020pressure}, we find that $c^{\cal R}_{xyxy} - c^0_{xyxy}=-\lambda p$ decreases linearly with pressure. Similar results are found for the other components of ${\hat C}^{\Delta}_{i,{\cal R}}-{\hat C}^{\Delta}_{i,0}$. 

\subsubsection{Probability distribution of local shear moduli ${\overline {\cal P}}(g^\ell)$}

We first show the probability distributions of the local shear moduli (i.e. $g^\ell \equiv c_{xyxy}$ from Eq.~\ref{eq:C}) from single Delaunay triangles in jammed disk packings (without performing rotations to the corresponding reference triangles) in Fig.~\ref{fig:pdfgtri}. We find two key features in ${\cal P}(g^\ell/{\cal S}_{g^\ell})$ for single Delaunay triangles. First, the probability of $g^\ell <0$ is small over the full range of $pN^2$. Second, since there are only five Delaunay triangle types in binary disk packings, ${\cal P}(g^\ell/{\cal S}_{g^\ell})$ displays multiple distinct peaks.  The peaks at large $g^\ell$ are maintained as $pN^2$ increases, but the peak at small $g^\ell$ decreases significantly.  Multiple peaks in ${\cal P}(g^\ell/{\cal S}_{g^\ell})$ are still found for $g^\ell$ based on subsystems composed of two adjacent triangles, whereas, ${\cal P}(g^\ell/{\cal S}_{g^\ell})$ possess a single peak for $g^\ell$ based on Voronoi polygons or larger subsystems, such as the square subsystems with side length $L/n$ and $n \le 12$.

\begin{figure}[t]
    \centering
    \includegraphics[width=0.425\textwidth]{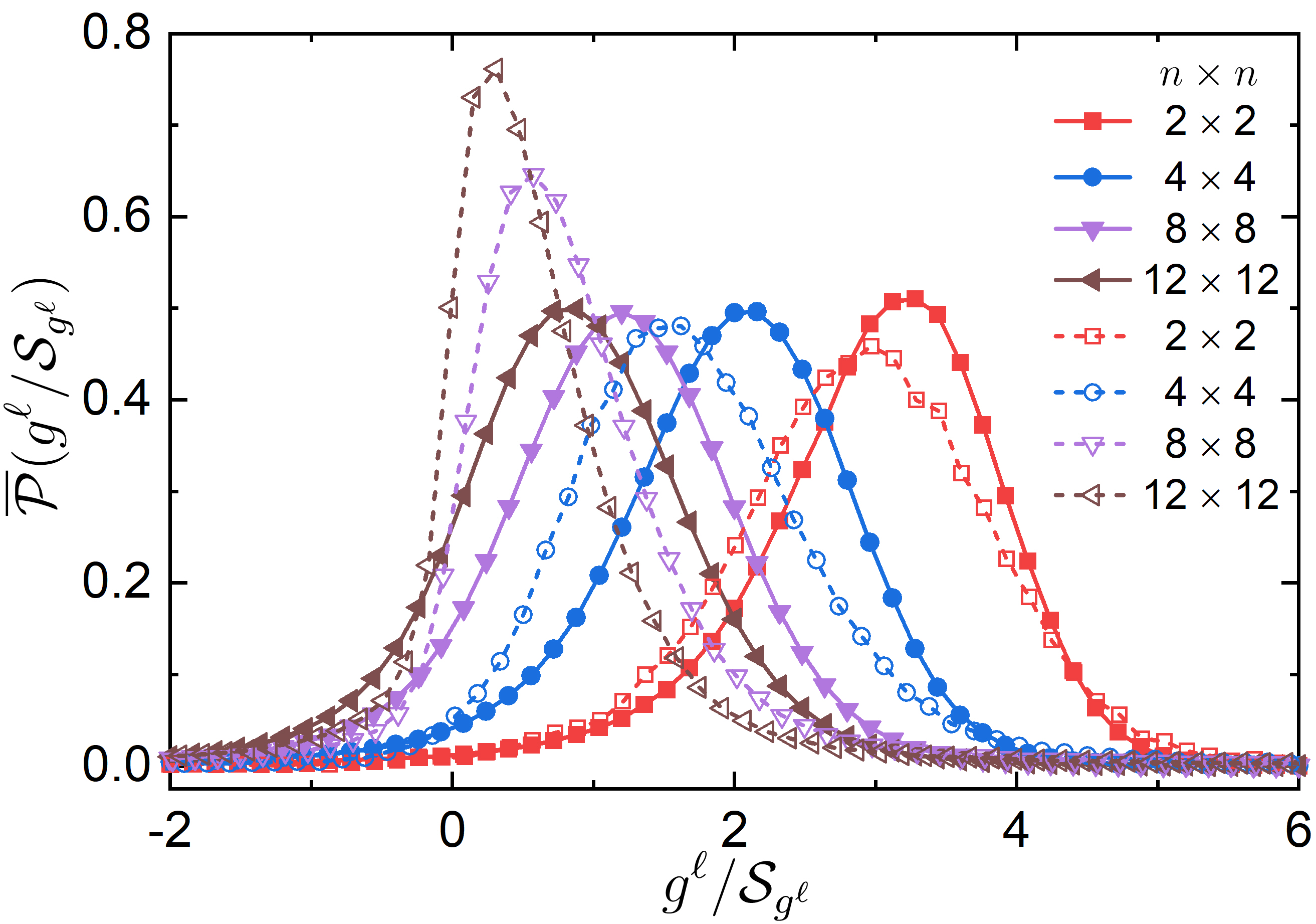}
\caption{Probability distributions of normalized local shear moduli ${\overline {\cal P}}(g^\ell/{\cal S}_{g^\ell})$ obtained using the affine-strain (filled symbols) and Delaunay triangulation (open symbols) methods at $pN^2\approx 10^4$ over a range of square subsystem sizes, $n^2=4$, $16$, $64$, and $144$.}
    \label{fig:pdfgcomp}
\end{figure}

Similar to ${\cal P}(g^\ell/{\cal S}_{g^\ell})$ obtained using the affine-strain method for calculating $g^\ell$, ${\cal P}(g^\ell/{\cal S}_{g^\ell})$ for the Delaunay triangulation method converges to ${\cal P}(G/{\cal S}_G)$ as the size of the subsystem increases (i.e. square subsystems with $n^2=1$). At large $pN^2$, ${\cal P}(g^\ell/{\cal S}_{g^\ell})$ is left-skewed with $\mu_3 < 0$ for the largest subsystem sizes and $\mu_3$ increases and becomes positive with decreasing subsystem size. In Fig.~\ref{fig:pdfgcomp}, we directly compare ${\cal P}(g^\ell/{\cal S}_{g^\ell})$ for local shear moduli calculated using the affine-strain and Delaunay triangulation methods in the $pN^2 \gg 1$ limit.  For small subsystems, e.g. $n^2 =64$ and $144$, ${\cal P}(g^\ell/{\cal S}_{g^\ell})$ for the two methods are significantly different. This result stems from the fact that the nonaffine contributions to the displacement fields play a more significant role in the mechanical response at smaller lengthscales ({\it cf.} Fig.~\ref{fig:disfield}). The affine-strain method for calculating the local shear modulus does not properly characterize the strain tensor of small subsystems, and thus does not accurately capture $g^\ell$. 

\begin{figure}[t]
    \centering
    \includegraphics[width=0.425\textwidth]{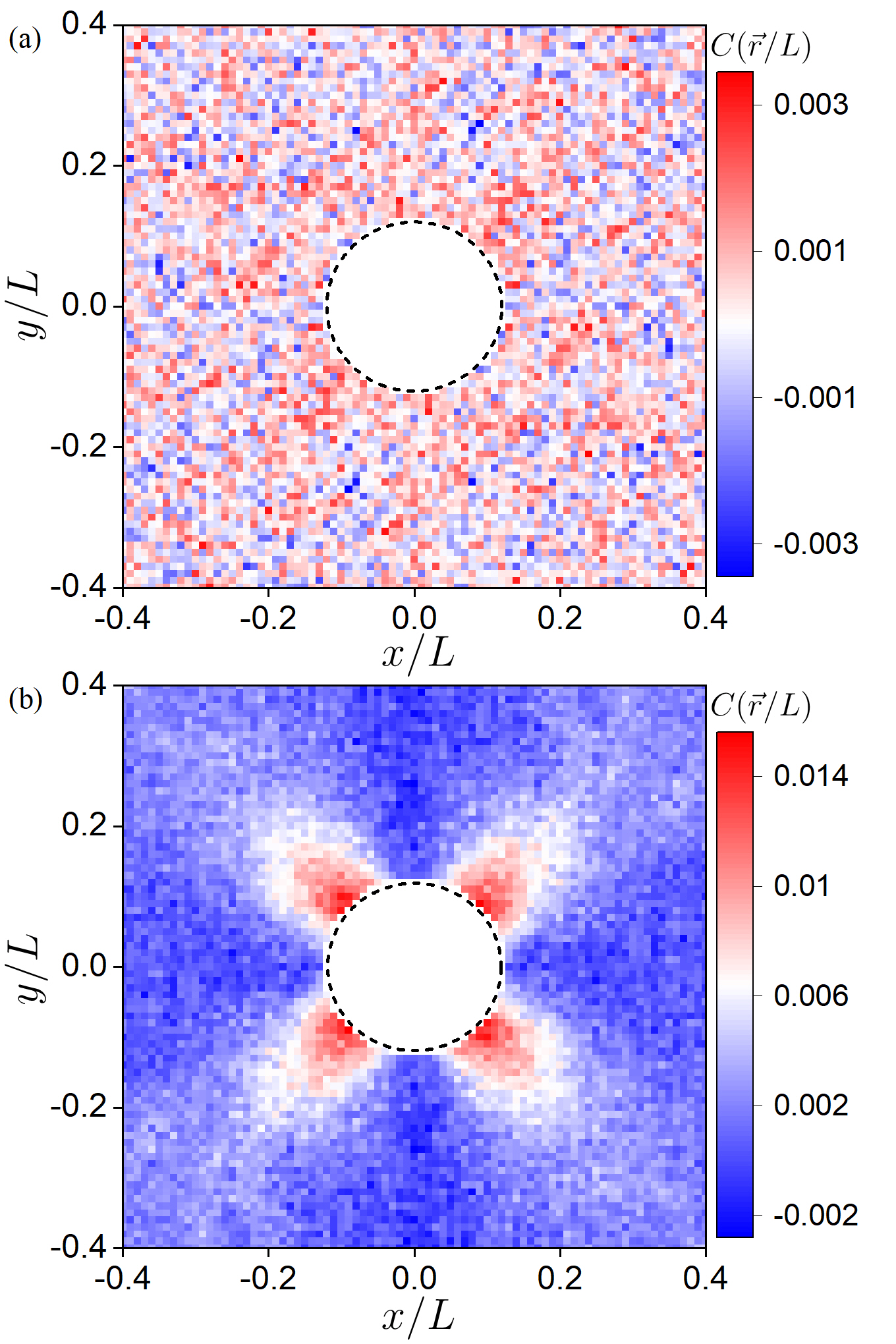}
\caption{Spatial correlation function $C({\vec r}/L)$ of the local shear moduli (obtained using the Delaunay triangulation method) of jammed disk packings using square subsystems with $n^2=144$ at (a) $pN^2=10^{-1}$ and (b) $10^{4}$. We do not display correlations for $r <\sqrt{2}L/n$ in the inner circular region.}
    \label{fig:corrg-tri}
\end{figure}

\begin{figure}[t]
    \centering
    \includegraphics[width=0.425\textwidth]{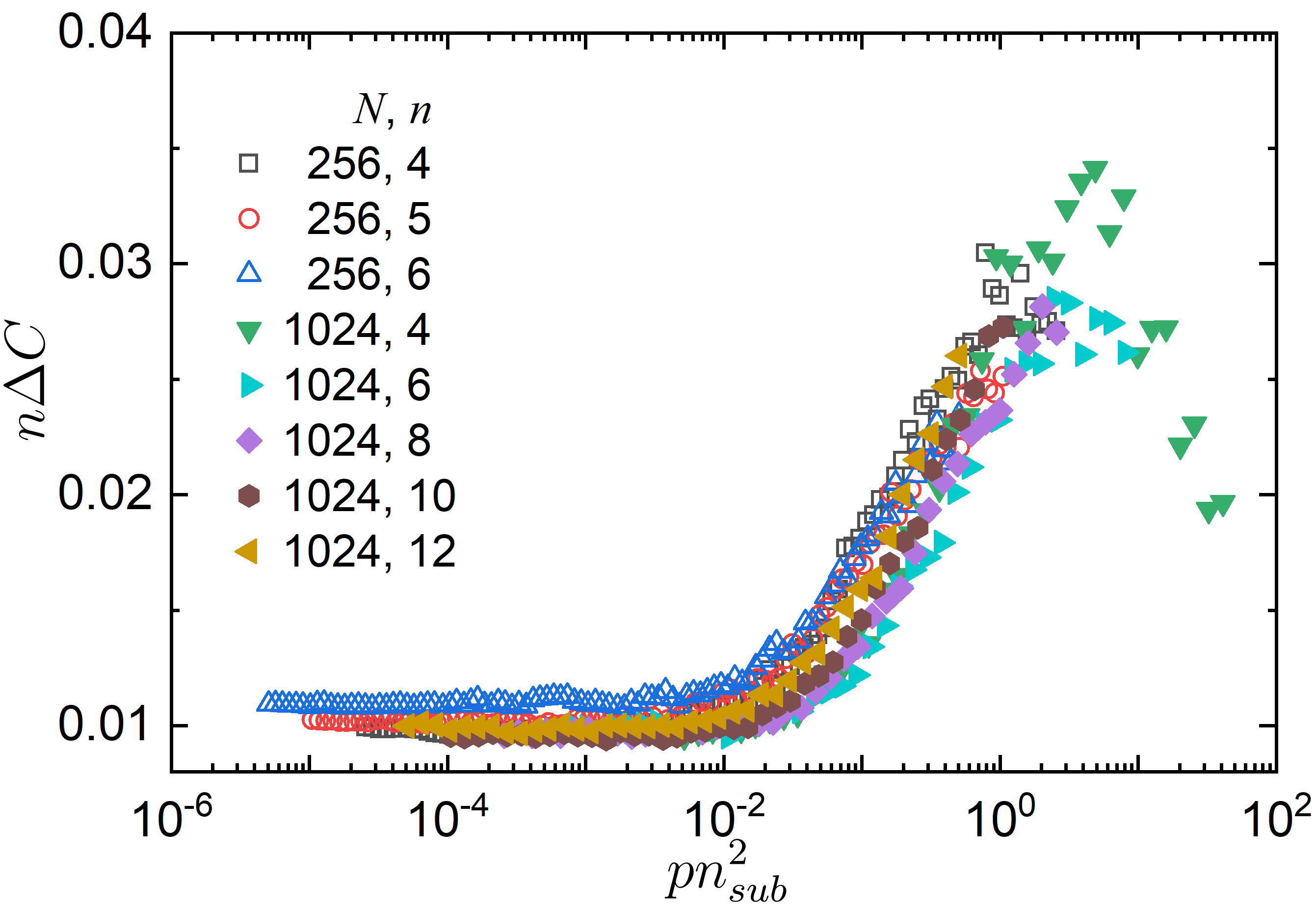}   
\caption{The standard deviation $\Delta C$ of the spatial correlation function of local shear moduli $g^\ell$ (multiplied by $n$) plotted as a function of $pn_{\rm sub}^2$, where $n_{\rm sub}=N/n^2$ is the average number of particles in each square subsystem with side length $L/n$. $g^\ell$ is calculated using the Delaunay triangulation method.}
    \label{fig:corrgtri-pn2}
\end{figure}

\begin{figure}[t]
    \centering
    \includegraphics[width=0.425\textwidth]{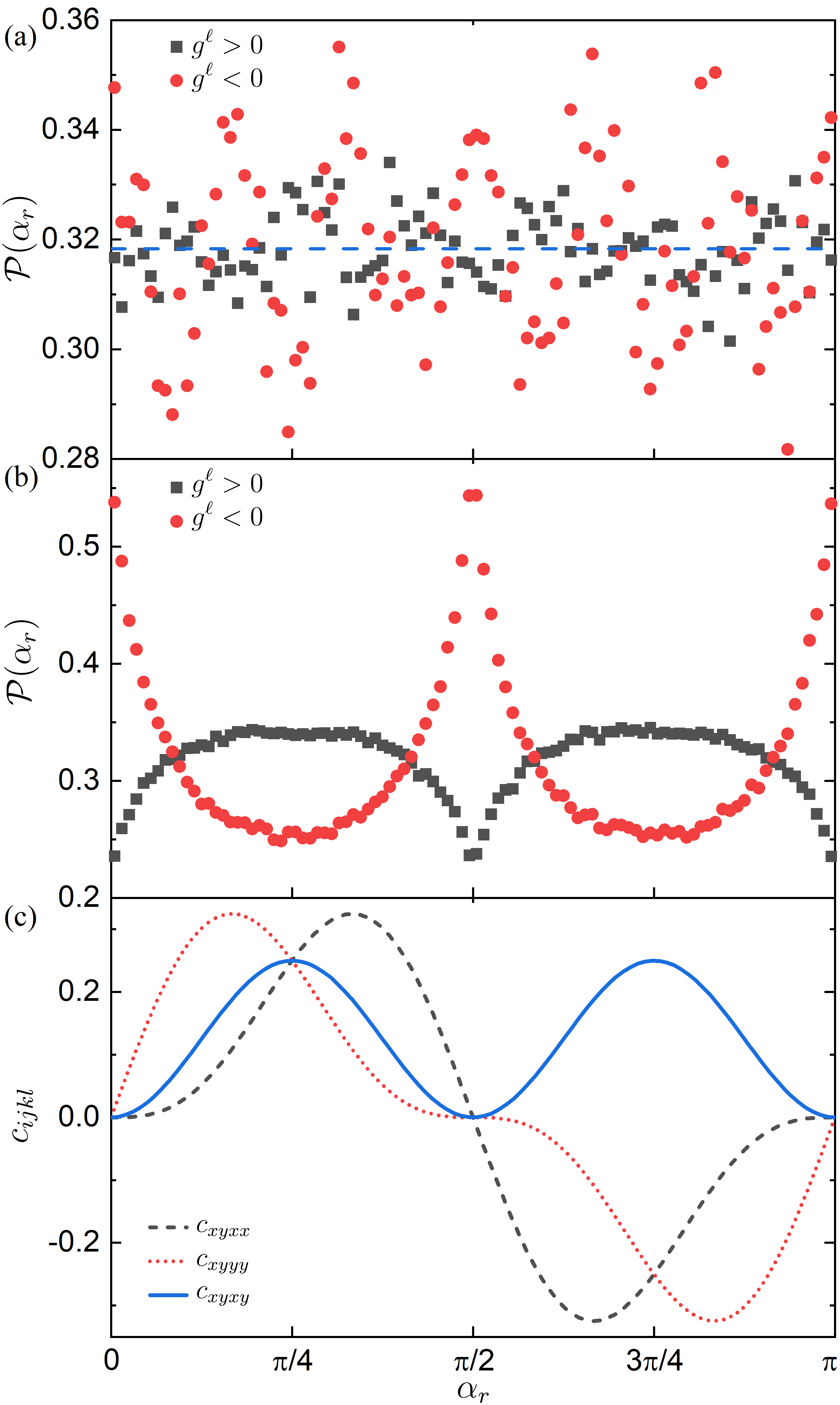}   
\caption{Probability distribution of the rotation angle ${\cal P}(\alpha_r)$ (Eq.~\ref{alpha}) that relates Delaunay triangles in jammed disk packings to the reference triangle types in Appendix~\ref{appxtristiff} for (a) type-3 and (b) type-1 triangles within subsystems composed of two adjacent triangles with positive (squares) and negative (circles) local shear moduli. (c) The $c_{xyxx}$, $c_{xyyy}$, and $c_{xyxy}$ components of the stiffness tensor (Eq.~\ref{eq:C}) for type-1 triangles as a function of $\alpha_r$. The horizontal dashed line in (a) corresponds to a uniform probability distribution over the range $0 \le \alpha_r \le \pi$.}
    \label{fig:pdfrot}
\end{figure}

\subsubsection{Spatial correlations of local shear moduli $g^\ell$}

In Fig.~\ref{fig:corrg-tri}, we show the spatial correlation function of the normalized and shifted local shear moduli, $C({\vec r})=\langle g^{\ell*}(0)g^{\ell*}(\vec{r})\rangle$, for $g^{\ell*}$ calculated using the Delaunay triangulation method. In contrast to $C({\vec r})$ for local shear moduli calculated using the affine-strain method, $C({\vec r})$ for local shear moduli calculated using the Delaunay triangulation method do not possess strong spatial correlations at low pressures, as shown in Fig.~\ref{fig:corrg-tri} (a).  
At high pressures, e.g. $p=10^{-2}$, $C({\vec r})$ regains long-range, four-fold symmetric spatial correlations, as shown in Fig.~\ref{fig:corrg-tri} (b). In Fig.~\ref{fig:corrgtri-pn2}, we show that the fluctuations in the spatial correlations, $n \Delta C$, collapse with $pn_{\rm sub}^2$, where $\Delta C = \sqrt{\langle (C-\langle C \rangle)^2 \rangle}$, $\langle \cdot \rangle$ indicates a spatial average, and $n_{\rm sub}=N/n^2$ is the average number of particles in each square subsystem with side length $L/n$. 
$n \Delta C \sim 0.01$ is constant in the low-pressure limit. When $pn_{\rm sub}^2 \gtrsim 10^{-2}$, $n \Delta C$ begins to increase, reaches a peak near $pn_{\rm sub}^2 \sim 1$, and then decreases for $pn_{\rm sub}^2 \gtrsim 1$. 
The low-pressure regime (i.e. $pn_{\rm sub}^2 \lesssim 10^{-2}$) for which $n \Delta C$ is constant corresponds to regime for which the spatial correlation function of the local shear moduli is short-ranged.

\subsubsection{Correlation between Delaunay triangle orientation and $g^\ell < 0$}

In Fig.~\ref{fig:pdfgtri}, we showed that the local shear moduli for single Delaunay triangles are nearly all positive over the full range of $pN^2$. However, we find that there are a significant number of negative local shear moduli for subsystems composed of two or more adjacent triangles even in the $pN^2 \gg 1$ limit. In this section, we investigate whether there is a difference in the orientation of the triangles within subsystems with positive versus negative local shear moduli. To address this question, we calculate the probability distribution of the rotation angle ${\cal P}(\alpha_r)$ of Delaunay triangles relative to the orientation of the reference triangle types in Appendix~\ref{appxtristiff}. As shown in Fig.~\ref{fig:pdfrot}, type-1 triangles in subsystems composed of two adjacent triangles with $g^\ell > 0$ are more likely to possess $\alpha_r \sim \pi/4$, which maximizes $c_{xyxy}$. In contrast, the most likely $\alpha_r$ for type-1 triangles within subsystems of two adjacent triangles with $g^\ell <0$ correspond to $\alpha_r$ that minimize $c_{xyxy}$. We find similar results for type-2, -4, and -5 triangles within subsystems composed of two adjacent triangles. However, for type-3 triangles, the stiffness tensor is independent of the rotation angle and thus ${\cal P}(\alpha_r) = 1/\pi$ is uniformly distributed between $0$ and $\pi$ for type-3 triangles within subsystems composed of two adjacent triangles and both positive and negative local shear moduli. (See Fig.~\ref{fig:pdfrot} (a).) 

\section{Conclusions and future directions}
\label{secsum}

In this article, we study the global and local shear moduli of jammed packings composed of $N$ repulsive, frictionless disks. The jammed disk packings are generated via isotropic compression at fixed boundary strain, and thus they can possess either positive and negative global shear moduli. We decomposed the ensemble-averaged global shear modulus into contributions from packings with positive and negative global shear moduli, $\langle G\rangle = (1-{\cal F}_-) \langle G_+\rangle + {\cal F}_-\langle G_-\rangle$, where ${\cal F}_-$ is the fraction of packings with negative global shear moduli and $\langle G_+\rangle$ and $\langle G_-\rangle$ are the ensemble-averaged values for packings with positive and negative global shear moduli, respectively. We find that $\langle G_+\rangle N$ and $\langle|G_-|\rangle N$ both scale as $\sim  (pN^2)^{1/2}$ for $pN^2 > 1$. Despite this, $\langle G\rangle N \sim (pN^2)^{\beta}$ with $\beta \gtrsim 0.5$ since ${\cal F}_-$ depends strongly on pressure~\cite{wang2021shear}. For $pN^2 < 1$, we find that $\langle|G_-|\rangle N \sim pN$ and $[\langle G_{+} \rangle-G_0/(1-{\cal F}_-)] N \sim (pN^2)^{1.33}$ possess different power-law scaling exponents. 

Not only do the ensemble-averaged global shear moduli scale with $pN^2$, but the probability distribution of global shear moduli ${\cal P}(G)$ collapses at fixed $pN^2$ and different values of $p$ and $N$.  We showed analytically that ${\cal P}(G)$ is a Gamma distribution with shape parameter $k=0.5$ in the $pN^2 \ll 1$ limit.  As $pN^2$ increases, ${\cal P}(G)$ transitions from a Gamma distribution with positive skewness in the small $pN^2$ limit to a skew-normal distribution with negative skewness in the large $pN^2$ limit. 

We also calculated the local shear moduli of jammed disk packings $g^\ell$ using two distinct methods: the affine-strain and Delaunay triangle methods. When using the affine-strain method, we find that ${\cal P}(G^*)$ and ${\cal P}(g^{\ell*})$ possess similar forms for $pN^2>1$ and the spatial correlation function of the local shear moduli $C({\vec r})$ is long-ranged with four-fold angular symmetry over the full range of $pN^2$. However, the affine-strain method does not accurately describe the strongly non-affine displacement fields that occur in response to applied deformations.  

In contrast, the spatial correlation function for $g^\ell$ calculated using the Delaunay triangulation method depends on $pn_{\rm sub}^2$, where $n_{\rm sub} =N/n^2$ is the number of disks per subsystem. In the $pn_{\rm sub}^2 \ll 1$ limit, the standard deviation of the spatial correlation function $n \Delta C \sim 0.01$ reaches a small plateau value and $C({\vec r})$ possesses weak spatial correlations.  $n \Delta C$ increases with $pn_{\rm sub}^2$ and $C({\vec r})$ begins to develop long-ranged, four-fold symmetric spatial correlations at $pn_{\rm sub}^2 >10^{-2}$.  We find very few single Delaunay triangles that possess $g^\ell <0$. However, there is an abundance of subsystems composed of two or more adjacent triangles that possess $g^\ell < 0$ and the individual triangles within these subsystems tend to orient in directions that minimize the components of the stiffness tensor. 

These results raise several important, open questions for future research. First,  what is the contribution of jammed packings with negative shear moduli to the ensemble-averaged density of vibrational modes $D(\omega)$?  Will the observed power-law scaling of $D(\omega)\sim\omega^4$ at low frequencies be affected by packings with negative shear moduli~\cite{lerner2016,kapteijns2018}?  Second, when we calculate the local shear moduli using Delaunay triangulation, we find that there are {\it growing} spatial correlations with {\it increasing} pressure $pn_{\rm sub}^2$ in contrast to previous work that shows growing spatial correlations with decreasing pressure associated with the isostatic length scale~\cite{wyart2005, heussinger2009, tighe2012}. What is the origin of the growing spatial correlations with increasing pressure? Third, the ratio $\langle G_a\rangle/\langle G_d \rangle \rightarrow 0$ in the $pN^2 \gg 1$ limit, and thus in this limit there are only two elastic moduli that characterize the mechanical response of jammed disk packings, i.e. $G_{d}\equiv G$ and the bulk modulus $B$. However, over a wide range of $pN^2$, both $G_a$ and $G_d$ (as well as $B$) are non-zero, and thus three elastic moduli characterize the mechanical response of jammed disk packings~\cite{baity2017}. Despite this, most previous work has focused on quantifying the pressure dependence of only two elastic moduli ($G$ and $B$) of jammed packings of spherical particles. In future work, we will characterize the pressure dependence of all non-trivial components of the stiffness tensor for jammed packings of spherical particles over the full range of $pN^2$.  Fourth, we will correlate regions with negative local shear moduli to ``soft spots''\cite{manning11,ding14,rainone20} and shear transformation zones~\cite{argon79,falk98,richard20} that occur during applied simple shear deformations~\cite{jin2021}.  Finally, we showed that the stiffness tensors vary with the different Delaunay triangle types for systems with short-ranged repulsive interactions, which is likely responsible for the anisotropic mechanical response for $pN^2 <1$. In future work, we will calculate the local shear moduli of amorphous packings with long-range attractive interactions, e.g. Lennard-Jones pairwise interactions. In this case, the stiffness tensors for the different Delaunay triangle types will likely be similar, which may shift the crossover from anisotropic to isotropic mechanical response to smaller pressures.

\section{Acknowledgments}
We acknowledge support from NSF Grant Nos. CMMI-1901959 (W.J. and C.S.O.), CBET-2002782 (W. J. and C. S. O.), DMREF-2118988 (C. S. O.), and CBET-2002797 (M.D.S.). This work was also supported by the High Performance Computing facilities operated by Yale’s Center for Research Computing.

\appendix
\section{Stiffness tensor of single Delaunay triangles}
\label{appxtristiff}

\begin{figure}[ht]
    \centering
    \includegraphics[width=0.425\textwidth]{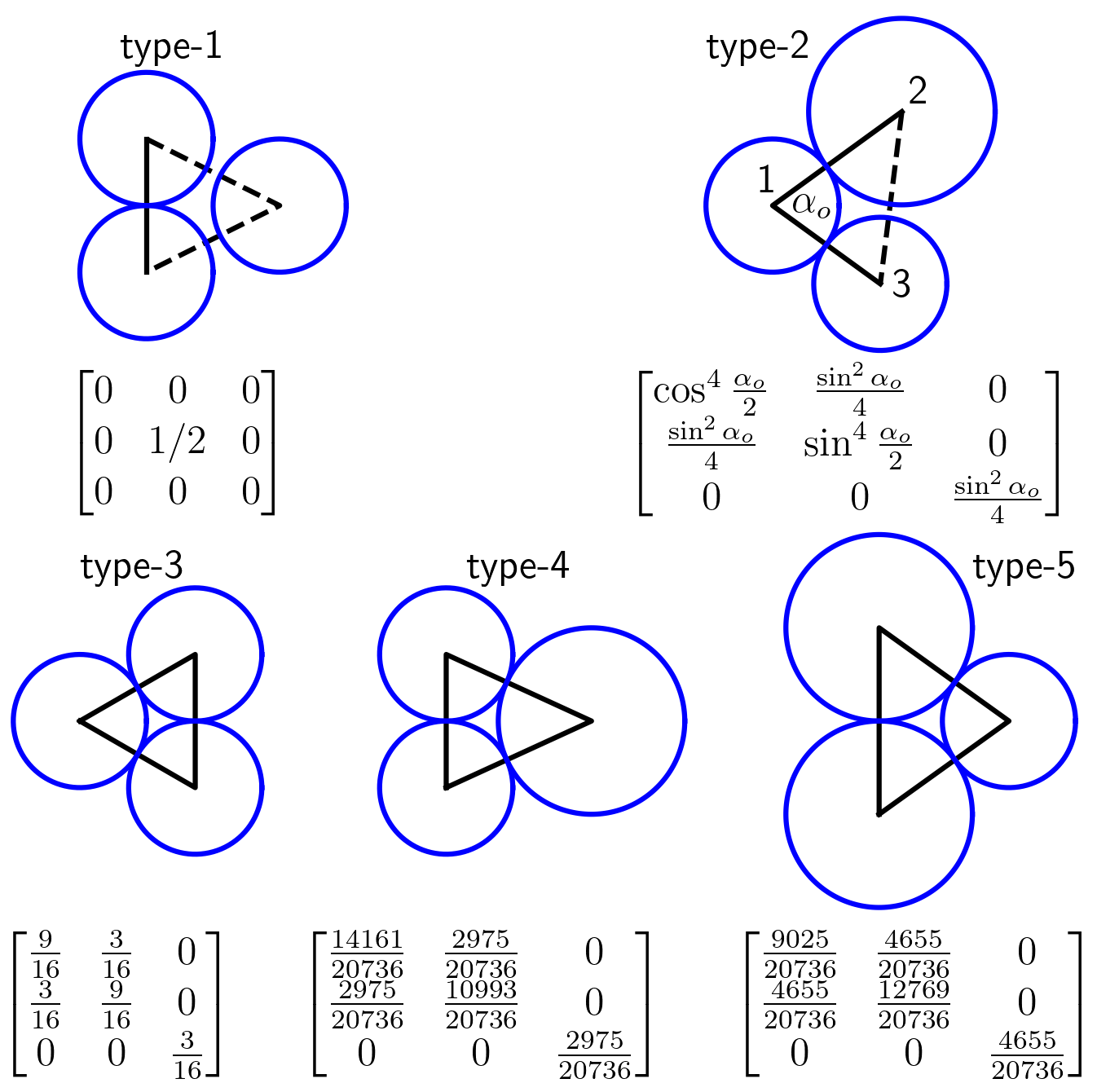}
\caption{Definitions of the five types of Delaunay triangles (with unique stiffness tensors) that occur in jammed packings of bidisperse disks. Solid lines indicate that adjacent disks are in contact, whereas dashed lines indicate that the disks are not in contact. For triange type-$2$, the angle $\alpha_o$ between ${\vec r}_{12}$ and ${\vec r}_{13}$ spans $\arccos\left(\frac{r_{12}^2+r_{13}^2-r_{23}^2}{2 r_{12}r_{13}}\right) < \alpha_o \lesssim 2.2$ rad. Below each triangle type, we display the corresponding stiffness tensors ${\hat C}_i^{\triangle}$ in the specific orientation shown. }
    \label{figtristiff}
\end{figure}

We define the five types of Delaunay triangles in Fig.~\ref{figtristiff}. Type-1 triangles possess a single contact among the three disks. This triangle type includes all possible disk size combinations since the stiffness matrix is the same for single-contact triangles with three large disks, three small disks, two small disks and one large disk, and two large disks and one small disk. The center-to-center separation vector for the two contacting disks is parallel to the vertical axis for the reference type-1 triangle. We define triangle type-2 as triangles with two contacts among any of the three disks since the stiffness matrix again does not depend on the size combinations. For the reference type-2 triangle, the horizontal axis bisects the angle $\alpha_o$ formed by the two segments between contacting disks. For triangle types-3, -4, and -5, all disks are in contact with each other and the center-to-center separation vector between the same-sized disks is parallel to the vertical axis.  For type-3, all three disks are the same size. For type-4, two of the disks are small and one disk is large. For type-5, two of the disks are large and one is small.  
The stiffness tensor ${\hat C}_{i0}^\Delta$ for the triangles with the reference orientation for each triangle type are provided in Fig.~\ref{figtristiff}. 

\section{Variation in the form of ${\cal P}(G)$ with $pN^2$}
\label{appxdistG}
\begin{figure}[ht]
    \centering
    \includegraphics[width=0.425\textwidth]{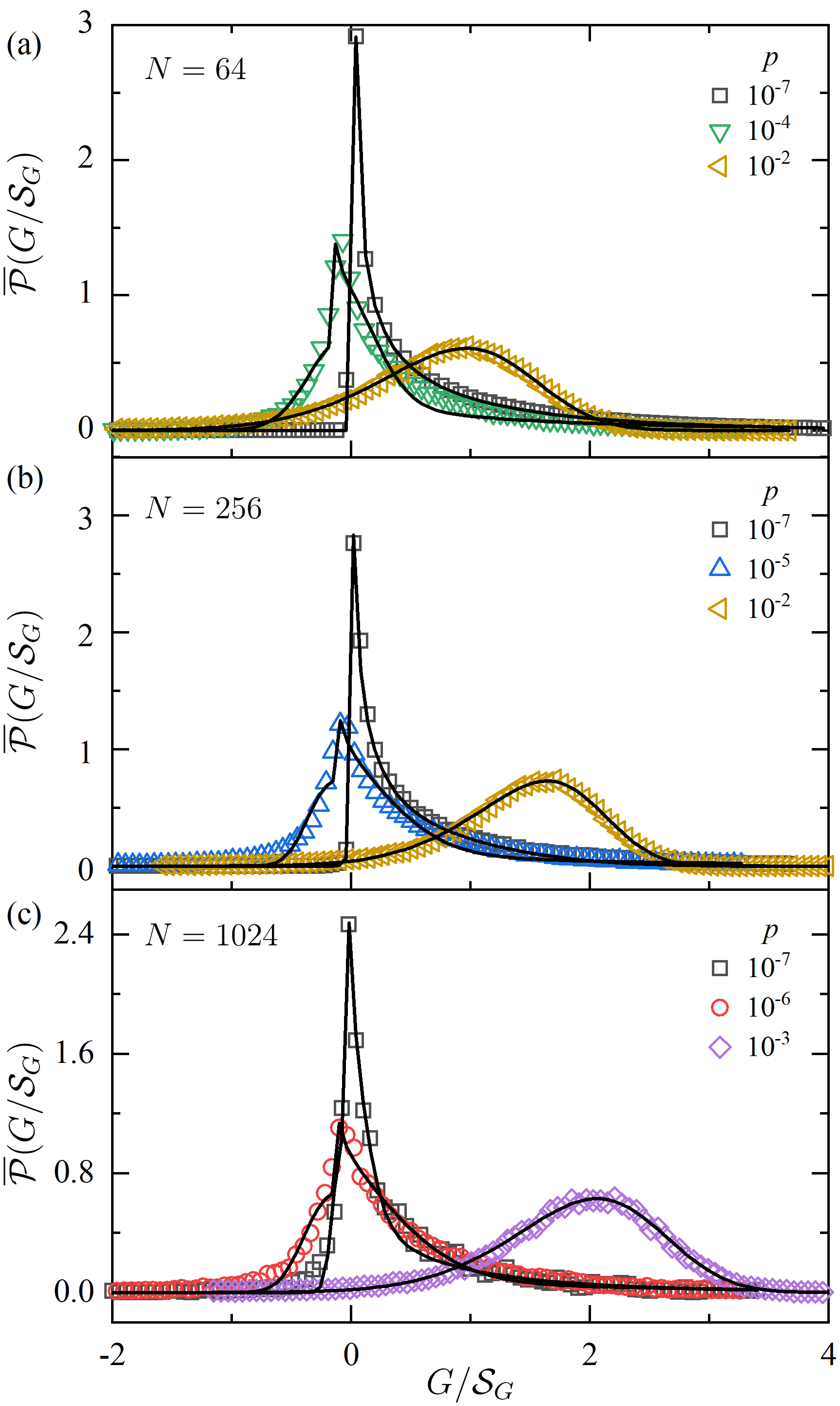}
\caption{The probability distribution of the global shear moduli $\overline{{\cal P}}(G/{\cal S}_G)$, where ${\cal S}_G$ is the standard deviation in $G$, for jammed disk packings over a range of pressures $10^{-7} \le p \le 10^{-2}$ and system sizes (a) $N=64$, (b) $256$, and (c) $1024$. The solid lines are examples of fits of ${\cal P}(G)$ using Eq.~\ref{eq:lerp}.}
    \label{fig:pdfG}
\end{figure}

In Fig.~\ref{fig:pdfG}, we show the probability distribution of the global shear moduli for jammed disk packings over a wide range of pressures $p$ and system sizes $N$.  In the $pN^2 \ll 1$ limit, ${\cal P}(G)$ obeys a Gamma distribution, which is right-skewed with ${\cal P}(G)=0$ for $G<0$,
\begin{equation}
\label{eq:gamma}
{\cal P}_{\Gamma}(G)=\frac{1}{\Gamma(k)\theta^k} G^{k-1} e^{-\frac{G}{\theta}},
\end{equation}
where $\Gamma(k)$ is the Gamma function, $k$ and $\theta$ are the shape and scale parameters, the mean is $\langle G\rangle=k\theta$, and the variance is ${\cal S}^2_G=k\theta^2$. Specifically, in Sec.~\ref{secglobal} we show that $k=1/2$ in the $pN^2 \ll 1$ limit, and thus Eq.~\ref{eq:gamma} can be rewritten as
\begin{equation}
\label{rewrite}
{\cal P}_{\Gamma}(G) = 2^{-1/4} \pi^{-1/2} {\cal S}_G^{-1} \left( \frac{G}{{\cal S}_G}\right)^{-1/2} e^{-\frac{G}{\sqrt{2}{\cal S}_G}}.
\end{equation}

As $pN^2$ increases, the peak in ${\cal P}(G)$ shifts to larger values of $G$, and the distribution evolves from a right-skewed Gamma distribution toward a left-skewed skew-normal distribution. In the $pN^2 \gg 1$ limit, we find that ${\cal P}(G) = {\cal P}_{SN}(G)$, where 
\begin{equation}
\begin{aligned}
    {\cal P}_{SN}(G)=\frac{2\sqrt{1-\frac{2\zeta^2}{\pi}}}{{\cal S}_G}\phi\left(\sqrt{1-\frac{2\zeta^2}{\pi}}\frac{G-\langle G\rangle}{{\cal S}_G}+\sqrt{\frac{2}{\pi}}\zeta\right)\\
    \Phi\left(\mu_{SN} \left(\sqrt{1-\frac{2\zeta^2}{\pi}}\frac{G-\langle G\rangle}{{\cal S}_G}+\sqrt{\frac{2}{\pi}}\zeta\right)\right),
    \label{eq:sn}
\end{aligned}
\end{equation}
$\zeta=\frac{\mu_{SN}}{\sqrt{1+\mu_{SN}^2}}$, $\mu_{SN}$ is the skew-normal shape parameter, 
\begin{equation}
\phi(x)=\frac{1}{2\pi}e^{-\frac{x^2}{2}},
\end{equation}
and
\begin{equation}
\Phi(x)=\frac{1}{2}\left[1+{\rm erf}\left(\frac{x}{\sqrt{2}}\right)\right].
\end{equation}

For intermediate values of $pN^2$, the form of ${\cal P}(G)$ can be approximated by a linear combination of ${\cal P}_{\Gamma}(G)$ and ${\cal P}_{SN}(G)$:
\begin{equation}
    {\cal P}(G)=s {\cal P}_{\Gamma}(G-G_m)+(1-s) {\cal P}_{SN}(G),
    \label{eq:lerp}
\end{equation}
where $0 \le s \le 1$, $G_m$ corresponds to the location of the maximum in ${\cal P}(G)$, and ${\cal P}_{\Gamma}(G) =0$ for $G<G_m$.

\section{Derivation of ${\cal P}(G)$ at jamming onset}
\label{appxPDF0}

In this Appendix, we include details of the derivation of the form of the probability distribution of the global shear moduli ${\cal P}(G)$ in the $pN^2 \ll 1$ limit.  As shown in Fig.~\ref{fig:GaGdSa} (b), the amplitude of the shear modulus is proportional to the square of the amplitude of the shear stress at jamming onset, $G_a= A_c \Sigma_a^2$ with proportionality constant $A_c = 10^{-5}$. Thus, Eq.~\ref{eq:GS} becomes 
\begin{equation}
G=2A_c\Sigma^2.
\label{eq:GSxyb}
\end{equation}
We first obtain the cumulative distribution function $F(G)$ and then calculate ${\cal P}(G)=dF/dG$. Since $G_a=G_d$ at jamming onset, $F(G)=0$ for $G<0$.  For $G\geq 0$, we have
\begin{equation}
\label{cumulative}
\begin{aligned}
F(G)&=\int_{-\sqrt{\frac{G}{2A_c}}}^{\sqrt{\frac{G}{2A_c}}} {\cal P}(\Sigma) \,d\Sigma\\
&=\text{erf}\left(\frac{\sqrt{G/A_c}}{2 \omega_s}\right),
\end{aligned}
\end{equation}
where ${\rm erf}(x)$ is the error function, using Eqs.~\ref{eq:pdfSxy} and~\ref{eq:GSxyb}.  The probability distribution is obtained by differentiating Eq.~\ref{cumulative} with respect to $G$:
\begin{equation}
{\cal P}_{\Gamma}(G)=\frac{1}{2 \omega_s\sqrt{\pi A_c G}}e^{-\frac{G}{4 A_c \omega_s^2}},
\end{equation}
which is a Gamma distribution with the shape parameter $k=0.5$.

\bibliography{refs.bib}

\end{document}